\documentclass[12pt]{article}
\pdfoutput=1
\usepackage{putex}
\usepackage{feyn}
\usepackage[vcentermath]{youngtab}

\usepackage{graphicx}
\usepackage{epstopdf}
\usepackage{enumerate}
\usepackage{cite}
\usepackage{tensor}
\usepackage{slashed}
\usepackage{feynmf}

\usepackage{hyperref}

\numberwithin{equation}{section}

\newcommand{\abs}[1]{\left\lvert #1 \right\rvert}

\newcommand {\be} {\begin {equation}}
\newcommand {\ee} {\end {equation}}

\newcommand {\bes} {\begin {equation*}}
\newcommand {\ees} {\end {equation*}}

\newcommand{\es}[2] {\begin{equation} \label{#1} \begin{split} #2 \end{split} \end{equation}}

\newcommand{\N}{\mathbb{N}}

\newcommand{\C}{\mathbb{C}}

\newcommand{\beq}{\begin{equation}}
\newcommand{\eeq}{\end{equation}}

\newcommand{\cF}{\mathcal{F}}

\newcommand{\p}{\partial}

\def\be{ \begin{equation} }
\def\ee{ \end{equation} }
\def\la#1{\label{#1}}

\def\half{{1\over  2}}

\def\vs{\vskip .1 in}

\def\subsec{\subsection}
\def\N{{\cal N}}
\def\A{{\cal A}}
\def\B{{\cal B}}
\def\C{{\cal C}}
\def\cF{{\cal F}}
\def\cH{{\cal H}}

\def\Vol{{\rm Vol}}
\def\IH{\mathbb{H}}
\def\IR{\mathbb{R}}
\def\eqr{\eqref}
\def\lab{\label}

\def\l{\lambda}
\def\t{\tau}
\def\lang{\langle}
\def\rang{\rangle}
\def\bul{$\bullet$\quad}

\def\Tr{\mathop{\rm Tr}}
\def\tr{\mathop{\rm tr}}

\newcommand{\bea}{\begin{eqnarray}}
\newcommand{\eea}{\end{eqnarray}}

\newcommand\lam{\lambda}
\newcommand\Lam{\Lambda}

\def\rar{\rightarrow}
\def\C{\mathcal{C}}

\def\eps{\epsilon}

\begin{document}

\institution{PU}{Department of Physics, Princeton University, Princeton, NJ 08544, USA}

\institution{CU}{DAMTP, Centre for Mathematical Sciences, University of Cambridge, CB3 0WA, UK}

\title{R\'enyi entropy, stationarity, and \\entanglement of the conformal scalar
}

\authors{Jeongseog Lee\worksat{\PU}, Aitor Lewkowycz\worksat{\PU}, Eric Perlmutter\worksat{\CU}, Benjamin R.~Safdi\worksat{\PU}
}

\abstract{
 
We extend previous work on the perturbative expansion of the R\'enyi entropy, $S_q$, around $q=1$ for a spherical entangling surface in a general CFT. Applied to conformal scalar fields in various spacetime dimensions, the results appear to conflict with the known conformal scalar R\'enyi entropies.  On the other hand, the perturbative results agree with known R\'enyi entropies in a variety of other theories, including theories of free fermions and vector fields and theories with Einstein gravity duals. We propose a resolution stemming from a careful consideration of boundary conditions near the entangling surface. This is equivalent to a proper treatment of total-derivative terms in the definition of the modular Hamiltonian. As a corollary, we are able to resolve an outstanding puzzle in the literature regarding the R\'enyi entropy of ${\cal N}=4$ super-Yang-Mills near $q=1$. A related puzzle regards the question of stationarity of the renormalized entanglement entropy (REE) across a circle for a (2+1)-dimensional massive scalar field.  We point out that the boundary contributions to the modular Hamiltonian shed light on the previously-observed non-stationarity.  Moreover, IR divergences appear in perturbation theory about the massless fixed point that inhibit our ability to reliably calculate the REE at small non-zero mass.  }

\date{}
\maketitle

\tableofcontents

\section{Introduction}

Quantum entanglement is a powerful probe of many-body quantum ground states and continuum quantum field theories (QFTs).  Given a reduced density matrix $\rho_\text{R}$, obtained by tracing $\rho = | 0 \rangle \langle 0 |$ over the degrees of freedom outside of a spatial entangling region, the R\'enyi entropies~\cite{renyi0,renyi1} 
\es{Renyi}{
S_q = {\log \tr \rho_\text{R}^q \over 1 - q} \,, \qquad q \geq 0 \,,
}
provide a natural method for quantifying entanglement.

The entanglement entropy (EE) (see~\cite{cardyCFT1,Eisert:2008ur,Nishioka:2009un} for reviews)
\es{ent}{
S = \lim_{q \to 1} S_q = - \tr \big(\rho_\text{R} \log \rho_\text{R}\big)
}
 across spherical entangling surfaces in flat spacetime has emerged as a particularly useful diagnostic of non-trivial critical behavior in quantum systems in various dimensions.  For example, the recently-discovered $F$-theorem~\cite{Jafferis:2011zi,Klebanov:2011gs,Myers:2010xs} states that the finite part of the EE across a circle in $(2+1)$-dimensional relativistic QFT satisfies a $C$-theorem.  More specifically, Casini and Huerta~\cite{Casini:2012ei} showed that the renormalized EE (REE)~\cite{Liu:2012eea} 
 \es{REE}{
 {\cal F}(R) = - S(R) + R S'(R) 
 }
 is a monotonically decreasing function of the radius $R$ of the entangling region.  At the conformal fixed points, ${\cal F}(R) = - F$, where $F = - \log Z_{S^3}$ is the finite part of the free energy of the theory conformally mapped to the three-sphere~\cite{ch2}.  
 
 The $F$-theorem is an exciting development given that there is an abundance of fixed points, many of which are believed to describe physical systems, in $(2+1)$ dimensions. In general spacetime dimensions, the EE across a spatial sphere is believed to be a measure of degrees of freedom in the same sense. In $(1+1)$ dimensions, the EE for an interval isolates the central charge $c$ at conformal fixed points, while in $(3+1)$ spacetime dimensions the EE across a two-sphere isolates the $a$ anomaly coefficient.  These quantities also satisfy $C$-theorems~\cite{Zamolodchikov:1986gt,Cardy:1988cwa,Komargodski:2011vj}.
  
While significant process in understanding the role of the EE across spherical entangling surfaces has been made in recent years, the relation of R\'enyi entropy to other aspects of QFT is less well understood.  Progress in understanding the CFT R\'enyi entropies across spherical entangling surfaces was made recently by expanding near $q = 1$~\cite{Perlmutter:2013gua}; the corrections to the EE in $d$ spacetime dimensions away from $q = 1$ are given by connected correlators of the Hamiltonian $H$ of the CFT conformally mapped to $\mathbb{R} \times \mathbb{H}^{d-1}$ at temperature $T = 1/(2 \pi R)$, where $\mathbb{H}^{d-1}$ is the  $(d-1)$-dimensional hyperbolic space of radius $R$. This is a general statement that follows solely from the definition of R\'enyi entropy. However, for a theory of free scalar fields in various dimensions, naive application of this expansion seems to conflict with taking derivatives of known results for $S_q$. 
 
 It is commonly assumed that the appropriate Hamiltonian $H$ on the hyperbolic space for computing quantum entropy is given by $H = \int_{\mathbb{H}^{d-1}} d^{d-1}x \sqrt{g} T_{\tau \tau}$, where $\tau$ is the time-like coordinate and $T_{\tau \tau}$ is the time-time component of the stress tensor.  In this work, we show that potential ambiguities in the definition of $T_{\tau\tau}$ familiar from ordinary QFT -- in particular, the presence or absence of total derivative terms -- must be properly accounted for. Applied to the theory of a free scalar field, these lessons eliminate the apparent discrepancy described above. 
 
Let us give slightly more detail. We compute the R\'enyi entropies via CFT partition functions on two different spacetimes. One is the hyperbolic cylinder, $ S^1 \times \mathbb{H}^{d-1}$, where the $S^1$ has circumference $2\pi R q$. In this frame, there is an apparent choice between conformal and non-conformal stress tensors that differ by a total derivative term. This is directly related to different ways of regularizing the boundary of $\IH^{d-1}$. The other spacetime is a conically-singular version of flat Minkowski space, ${\cal C}_q\times \IR^{d-2}$, with a deficit angle $2\pi(q-1)$ along the entangling surface; this is the spacetime generated directly by the replica trick. In this frame, there are different ways of regularizing the conical singularity that may differ by boundary terms. To compute the conventional R\'enyi entropies, one regulates the conical singularity by putting in a hard cut-off a distance $\epsilon$ away from the entangling surface.  From this point of view, boundary terms appear in the modular Hamiltonian localized a short distance away from the entangling surface.  

We show the necessity of including these boundary terms through explicit computations. Using the connected three-point function of $H$ built from the {\it conformal} stress tensor, we derive $S''_{q=1}$ for a general CFT, where the primes denote derivatives with respect to $q$. We perform the calculation by utilizing a conformal mapping back to $\IR^d$, where the stress tensor three-point function is fixed by conformal symmetry up to three calculable, theory-dependent constants. We then check our result for $S''_{q=1}$ against derivatives of previous results for $S_q$ in a handful of CFTs. Applied to CFTs with Einstein gravity duals and to free Dirac fermions across dimensions, and to free vector fields in $d = 4$, we find perfect agreement; applied to the conformal scalar, we find an apparent mismatch instead. 

The resolution to this issue was laid out above: in taking derivatives of $S_q$ with respect to $q$, one must take into account boundary contributions from the singular cone. We substantiate this statement with three different (matching) calculations of $S''_{q=1}$, performed on the spaces ${\cal C}_q \times \mathbb{R}^{d-2}$, $S^d$, and on $ S^1 \times \mathbb{H}^{d-1}$.   We also apply these conclusions to the computation of $S''_{q=1}$ in $\N=4$ super-Yang-Mills. Due to the boundary terms in the stress tensor, this quantity depends on the `t Hooft coupling $\lambda$, consistent with known results at weak and strong coupling.  If one were to neglect the boundary terms, non-renormalization of the central charges would seem to incorrectly imply non-renormalization of $S''_{q=1}$ instead. 

These stress tensor subtleties also come to bear on the question of stationarity of REE under relevant deformations of a CFT. A naive application of conformal perturbation theory to the definition of REE implies stationarity, but this conflicts with lattice computations of the REE for a massive free scalar of mass $m$ and with holographic computations~\cite{Klebanov:2012va,Nishioka:2014kpa}. Part of the resolution for the free scalar is due to the necessity of including the boundary contributions to the modular Hamiltonian; in this case, the term linear in $m^2$ in the conformal perturbation theory does not vanish.  Without the boundary contributions, the linear term would vanish identically.

 There is another aspect to this issue, however. We argue that the perturbation theory itself is not well defined due to IR divergences.  The IR divergences relate terms at different orders in the perturbative expansion. 

While the match to previously calculated results privileges the singular over the regularized cone, in Sec. 5 we comment further on the physics of the regularized cone and its possible role in alternative definitions of R\'enyi entropy. 

\vs
{\bf Note added:}  While this work was in its final stages of preparations, \cite{Hung:2014npa} appeared, which overlaps with parts of our Section 3. Where they overlap, our results agree; \cite{Hung:2014npa} notes, but does not address, the subtleties associated with the conformal scalar R\'enyi entropy.

 \section{R\'enyi entropy and the modular Hamiltonian} \label{sec: 2}

We begin by considering CFT R\'enyi entropies across spherical entangling surfaces in $d$ flat spacetime dimensions.  These quantities may be calculated from the thermal free energy on the hyperbolic space $\mathbb{H}^{d-1}$, where the temperature is $T = 1 / (2 \pi R q)$, and $R$ is the radius of $\mathbb{H}^{d-1}$, which we subsequently set to unity~\cite{Hung:2011nu,ch2}.  
The thermal partition function may be calculated by Wick rotating and considering the space ${\cal H}_q^d = S^1 \times \mathbb{H}^{d-1}$ with compactified Euclidean time: 
\es{}{
ds_{H_q^d}^2 = d\tau^2 + du^2 + \sinh^2u \,d \Omega_{d-2}^2 \,, \qquad \tau \sim \tau + 2 \pi q \,.
}
When $q=1$, we define ${\cal H}^d \equiv {\cal H}^d_{q=1}$.
Defining ${\cal F}_q = - \log Z_q$, with $Z_q$ the Euclidean partition function on ${\cal H}_q^d$, the R\'enyi entropy is given by 
\es{}{
S_q = {q {\cal F}_1 - {\cal F}_q \over 1 - q} \,.
}

The partition function on ${\cal H}^d_q$ is naturally written as 
\es{Zq}{
Z_q = \tr\left( e^{- 2 \pi q H_\tau} \right) \,,
}
where $H_\tau$ is the Hamiltonian that generates translations along the $S^1$.  Using this relation, it was pointed out in~\cite{Perlmutter:2013gua} that derivatives of $S_q$ with respect to $q$ generate connected correlation functions of $H_\tau$.  In particular, expanding $S_q$ in the vicinity of $q = 1$ leads to 
\es{Expandq}{
S_q  = S_1 + 2 \pi \sum_{n = 1}^\infty {1 \over (n+1)!} \left.\partial_q^n E_q \right|_{q = 1}(q-1)^n \,, \qquad S_1 = - {\cal F}_1 + E_1 \,,
} 
where $E_q = \langle H_\tau \rangle_q$.  The subscript $q$ is a reminder that the expectation value is to be computed at inverse temperature $2 \pi q$.  We may further simplify~\eqref{Expandq} by writing
\es{En}{
\left. \partial_q^n E_q \right|_{q = 1} =  (-1)^n (2 \pi)^n \langle \underbrace{H_\tau H_\tau \cdots H_\tau}_{n+1} \rangle_{q=1}^\text{conn}   \,.
}

The Hamiltonian $H_\tau$ is simply related to the stress tensor, $H_\tau = \int_{\mathbb{H}^{d-1}} d^{d-1}x \sqrt{g} T_{\tau \tau}$.  
Because the field theory is conformally invariant, it is natural to assume that the stress tensor is the conformal one. We will show that this statement depends on the boundary conditions at infinity. One particularly important exceptional theory is that of the free conformally coupled scalar field. For the scalar to be conformal, we need to add a conformal mass term to the Lagrangian:  ${\cal L} \supset {d - 2 \over 8 (d-1)} \sqrt{g} {\cal R} \phi^2$.  Note that on ${\cal H}^d_q$ the curvature scalar is given by ${\cal R}=-(d-2)(d-1)$. This term contributes to the stress tensor in two ways:  (1) there is a contribution to $T_{\tau \tau}$ that comes from varying $\sqrt{g}$, and (2) there is a contribution that arises from varying ${\cal R}$.  The second contribution leads to $T_{\tau \tau} \supset \nabla^2 \phi^2$, where the Laplacian is only over the coordinates on $\mathbb{H}^{d-1}$. It is important to remember that in deriving this term, one must integrate by parts on $\mathbb{H}^{d-1}$.  Here, the boundary conditions on the hyperbolic space become important. 

Following the normal procedure leads to the conformal stress tensor\footnote{Note that since the metric is a direct product of $S^1 \times \mathbb{H}^{d-1}$, ${\cal R}_{\tau \tau}=0$}
\es{confScalar}{
T_{\tau \tau}^\text{conf} = (\partial_\tau \phi)^2 - {1 \over 2} \partial_\mu \phi \partial^\mu \phi + {(d - 2) \over 8 (d-1)} {\cal R}  \phi^2 + {d - 2 \over 4 (d-1)} \nabla^2 \phi^2 \,.
}
We will show that for the scalar theory, when we regularize the space ${\cal H}^d_q$ by putting a cutoff at infinity, it is in fact what we call the ``non-conformal'' stress tensor  
\es{unimpScalar}{
T_{\tau \tau} = (\partial_\tau \phi)^2 - {1 \over 2} \partial_\mu \phi \partial^\mu \phi + {(d - 2) \over 8 (d-1)} {\cal R}   \phi^2
}
that enters into $H_\tau$.\footnote{Due to the presence of the curvature term in \eqr{unimpScalar}, this is not the same thing as the ``unimproved'' stress tensor.} This stress tensor can be thought of as arising from not integrating the variation of the curvature scalar by parts.  

Because these two stress tensors differ only by a total derivative, it may appear that this difference does not affect $H_\tau$.  This is in fact not the case, since we are introducing a boundary at some large value of the $u$ coordinate, which we call $u_0$~\cite{ch2}. Note that $u_0$ is related to the UV cutoff $\epsilon$ of the theory through the relation $\epsilon \sim e^{- u_0}$.  In particular, the regularized volume of the hyperbolic space is found by subtracting off the power-law divergences in $\epsilon$ from the integral 
\es{}{
\Vol({\mathbb{H}^{d-1}}) = \Vol( S^{d-2} ) \int_0^{u_0} du \sinh^{d-2} u \,.
}    
This leads to the result~\cite{Casini:2010kt,Diaz:2007an,ch2}
\es{hypvolume}{
\Vol({\mathbb{H}^{d-1}}) = (-1)^{\left\lfloor \frac{d}{2}\right\rfloor } \frac{\pi^{\frac{d-2}{2}}}{\Gamma (\frac{d}{2})}
\left\{ 
\begin{array}{cc}
\pi \,, & \text{$d$ odd} \\
-2 \log (R/\epsilon) \,, & \text{$d$ even} \,.
\end{array}
\right.
}
Note that in even-dimensional theories, the $\log \epsilon$ dependence of the R\'enyi entropy arises through the regularized volume of $\mathbb{H}^{d-1}$.

\subsection{Boundary conditions and entanglement: singular vs. regularized cone }

For certain theories, there is an ambiguity in the definition of the EE. In the replica trick method~\cite{Callan199455}, the EE is defined through a partition function on a conically singular manifold. There is infinite curvature concentrated at the tip that needs to be regulated.  The ambiguity arises from the fact that one can do so in (at least) two different ways: we refer to these as the ``singular cone'' and the ``regularized cone''. This is similar in spirit to the ambiguities that arise when trying to separate the Hilbert spaces of gauge theories in the lattice (see \cite{Casini:2013rba} and references therein). 

In the singular cone, one places a hard-wall at a fixed distance $\epsilon$ from the singularity~\cite{Callan199455}. Note that this space has ${\cal R}=0$ and a boundary. In the regularized cone, we can instead smooth out the singularity as we get closer to the tip \cite{Fursaev:1995ef}. To do this, one changes the metric such that near the tip the space is locally flat, but far from the tip the metric is unchanged. The regularized cone has ${\cal R}^{reg}(q) \not=0$ and no boundary.

The singular cone and the regularized cone have natural interpretations in terms of how we treat the boundary of the space ${\cal H}^d_q$\cite{Lewkowycz:2013laa}.  In the first case, we simply cut the space off at some large value for the $u$ coordinate $u_0$.  In the second case, we take $T^{-1} = 2 \pi q$ at small $u$, but then we ``regulate'' the spacetime so that at large $u$, the inverse temperature reverts to $T^{-1} = 2 \pi$.  In the first case, the curvature scalar is simply ${\cal R}=-(d-2)(d-1)$, independent of $q$, while in the latter case the curvature receives $q$-dependent corrections ${\cal R}^{reg}(q)$ at large $u$.\footnote{Note that if we were working in Rindler space, which is conformally equivalent to ${\cal H}_q^d$~\cite{ch2}, the regularized cone method maps to the regularized manifold described in \cite{Fursaev:1995ef}, which does not have a boundary.}

In many situations, the method used to regulate the cone/${\cal H}^d_q$/Rindler space does not affect physical quantities.  However, the choice of regularization {\it does} affect theories for which ${\delta S / \delta \partial g} \not =0 $. In these cases, one must integrate by parts to derive the stress tensor.  If we are using the singular cone method, we must be careful when integrating by parts, since the spaces have boundaries.

This discussion is reminiscent of computations of quantum black hole entropy via the conical singularity method. In \cite{Kabat:1995eq}, the contribution of quantum fields to black hole entropy was considered. The usual way to compute black hole entropy  \cite{Gibbons:1976ue} consists of evaluating the gravitational action as a function of the temperature and then taking derivatives with respect to the temperature. This prescription, in which we have a family of smooth geometries labeled by the temperature, can be thought of as being analogous to the regularized cone. When computing the quantum contribution to the black hole entropy from conformal scalars near a black hole (and trying to express it as EE), there is a ``contact term'' on the horizon. This can be understood as the Wald entropy due to the quantum fields, coming from their direct coupling to curvature of the regularized cone: in other words, $S_{BH}=S_{EE}+\langle S_{wald} \rangle$.  This type of term also appears in the contribution of bulk quantum fields to $1/N$-corrected holographic EE  \cite{Ryu:2006ef,Faulkner:2013ana}.  In this case, the prescription is to compute bulk EE across the Ryu-Takayanagi surface, supplemented by possible Wald-type terms.

Returning to R\'enyi entropy, the previous discussion makes clear that the definition of the Hamiltonian $H_{\t}$ is sensitive to the choice of regularization. In the scalar theory, working on the singular cone gives the non-conformal stress tensor $T_{\t\t}$, while the regularized cone gives the conformal stress tensor $T_{\t\t}^{\rm conf}$, as will be discussed. This is because if we are in the regularized cone, we can integrate by parts without problems, while additional boundary terms arise when integrating by parts in the singular cone~\cite{Lewkowycz:2013laa}. 

With regard to computing EE and R\'enyi entropy, one might be tempted to ask, ``which regularization of the cone should we use?'' The singular cone is the regularization appropriate for the conventional definition of EE. For example, we will see that the singular cone method matches the results of lattice calculations of R\'enyi and EE as well as previous analytic calculations of these quantities, while the regularized cone method does not. Accordingly, in taking $q$-derivatives as in \eqr{Expandq}, we should be using the non-conformal stress tensor in our definition of $H_{\t}$. This is one punchline of this paper. 

We also wish to point out that while the singular cone reproduces established results for R\'enyi entropy, the regularized cone is in a sense the more natural background in which to study conformal field theory. We postpone further comment on this perspective to Sec. 5.

\subsection{Warmup: stationarity on $S^1 \times \mathbb{H}^2$}
\label{sec: statE}

There is a straightforward example that illustrates the fact that it is the non-conformal stress tensor $T_{\tau \tau}$ that enters into the Hamiltonian for the scalar field.  We consider the connected two-point function
\es{warmup_stat}{
\int_{{\cal H}^3} d^3x \sqrt{g(x)} \langle H_\tau  \phi^2(x) \rangle_{q=1}^\text{conn} =  \int_{{\cal H}^3} d^3x \sqrt{g(x)} \int_{\mathbb{ H}^2 } d^2y \sqrt{g(y)} \langle T_{\t\t}(y)  \phi^2(x) \rangle_{q=1}^\text{conn}\,.
}

  If $T_{\tau \tau}$ were the conformal stress tensor, then this quantity would vanish, since 
  \es{conf0}{
  \langle T^\text{conf}_{\tau \tau}(y) \phi^2(x) \rangle_{q=1} =0 = \lang  \phi^2(x) \rang_{q=1}  \,.
  }
    This follows from the fact that ${\cal H}^3$ is related to flat space by a conformal transformation, and the two-point function of the conformal stress tensor with a primary operator in flat space vanishes, along with the one-point function of the primary operator. This argument is used in \cite{Rosenhaus:2014woa} in the general context of perturbations of EE by the addition of relevant operators, but as we see here, the situation is more subtle. 
  
  We may calculate the quantity in~\eqref{warmup_stat} by calculating the partition function of the massive scalar field on ${\cal H}^3_q$.  The action of the theory is given by 
  \es{actionScalar}{
  I = \int_{{\cal H}^3_q} d^3x \sqrt{g} \left[ {1 \over 2} (\partial_\mu \phi)^2 - {1 \over 8} \phi^2 + {m^2 \over 2} \phi^2 \right] \,,
  } 
  so that when $m^2 = 0$ the theory is conformal.  The Euclidean free energy ${\cal F}_q(m^2)$ of this theory was calculated as a function of $m^2$ and $q$ in~\cite{Klebanov:2011uf}:
  \es{}{
  {\cal F}_q(m^2) = \int_0^\infty d \lambda {\cal D}(\lambda) \left[ \log \left( 1 - e^{-2 \pi q \sqrt{\lambda + m^2 } }\right) + \pi q \sqrt{\lambda + m^2 } \right] \,,
  }
  with the density of states given by 
  \es{}{
  {\cal D}(\lambda) d \lambda = {\Vol(\mathbb{H}^2) \over 4 \pi } \tanh(\pi \sqrt{\lambda}) d \lambda \,.
  }
  Note that $\lambda$ parameterizes the eigenvalues of the $m^2 = 0$ Laplacian on $\mathbb{H}^2$.
  
  The two-point function in~\eqref{warmup_stat} is related to ${\cal F}_q(m^2)$ through the equation 
  \es{215}{
\int_{{\cal H}^3} d^3x \sqrt{g(x)} \langle H_\tau  \phi^2(x) \rangle_{q=1}^\text{conn} =  - {1 \over \pi} \left. \partial_q \partial_{m^2} {\cal F}_q(m^2) \right|_{q = 1, m^2 = 0} \,.  
}
The resulting integral is UV divergent, but we may regularize the integral through a cut-off in $\lambda$ $\sim 1 / \tilde \epsilon^2$.  The scaling follows from the fact that $\lambda$ has mass dimension two.  Then, 
 \es{statH3}{
\int_{{\cal H}^3} d^3x \sqrt{g(x)} \langle H_\tau  \phi^2(x) \rangle_{q=1}^\text{conn} &= {1 \over 4 } \int_0^{1/ \tilde{\epsilon}} d\lambda { 1 \over \sqrt{\lambda}} - \pi \int_0^\infty d \lambda \csch(2 \pi \sqrt{\lambda}) \\
&= {1 \over 2 \tilde{\epsilon}} - {\pi \over 16} \\
}
The fact that the finite, $\tilde \epsilon$-independent term above is non-vanishing tells us that $H_\tau$ is computed from the non-conformal $T_{\tau \tau}$.

We may understand this fact is a somewhat more illuminating way by an explicit computation of $\langle H_\tau \int \phi^2 \rangle_{q=1}^\text{conn}$.
Because of~\eqref{conf0}, we only have to compute $\langle ( \nabla^2 \phi^2 )\int \phi^2 \rangle_{q=1}^\text{conn}$:
\es{ex1b}{
\int_{{\cal H}^3} d^3x \sqrt{g(x)} \langle H_\tau  \phi^2(x) \rangle^\text{conn}_{q=1} & = 
-\frac{1}{8} \text{Vol}({\mathbb H}^2) \int_0^\infty du \sinh u \\
&\qquad \int_0^{2 \pi} d\phi_0 \int_0^{2 \pi} d\tau  \langle \nabla^2 \phi^2(0) \phi^2(u,\phi_0,\tau )\rangle^\text{conn}_{q=1} \\ &=-\int_0^\infty du \sinh u\int_0^{2 \pi} d\tau  \frac{\cos \tau  \cosh u-1}{16 (\cos \tau -\cosh u)^3} \\
&=-\frac{\pi}{16} \,.
}
where $\phi_0$ is the $\IH^2$ angular coordinate. Note that we have used translation invariance to place the first $\phi^2$ at the origin and factor out a regulated volume of the hyperbolic space.  The regulated volume also introduces a UV divergent term, but the important observation is that the finite term above matches that in~\eqref{statH3}.
In computing~\eqref{ex1b}, we explicitly used the propagator for the scalar field on ${\cal H}^3$:
\es{}{
&\langle \phi(u_1, \phi_1, \tau_1) \phi(u_2, \phi_2, \tau_2) \rangle_{q = 1} = \\
&{1 \over 4 \pi \sqrt{ 2\cosh(u_1) \cosh(u_2) -2 \cos(\phi_1 - \phi_2) \sinh(u_1) \sinh(u_2) - 2\cos(\tau_1 - \tau_2) }} \,.
}

An alternative way to evaluate~\eqref{215} is to calculate $ \langle \phi^2 \rangle_q$ at finite $q$ in the $m^2 = 0$ CFT.  The calculation of  $ \langle \phi^2 \rangle_q$ was performed in~\cite{Cardy:2013nua}, where it was found that $4 \pi \langle \phi^2 \rangle_q=-(q-1) \frac{\pi}{8}$.  Then, a straightforward calculation leads again to the result in~\eqref{statH3}:
\es{}{
- {1 \over \pi} \left. \partial_q \partial_{m^2} {\cal F}_q(m^2) \right|_{q = 1, m^2 = 0}=-\text{Vol}({\mathbb H}^2) \partial_q \langle \phi^2 \rangle_q=-\frac{\pi}{16} \,.
 }

 \section{The R\'enyi entropy near $q = 1$}
 
 In this section we calculate the R\'enyi entropy perturbatively in $q$, near $q = 1$, using the formalism of Sec.~\ref{sec: 2}.  We begin by discussing $ S'_{q=1}$ and then we discuss $S''_{q = 1}$.  Throughout this section, we will work in an arbitrary spacetime dimension $d$, unless otherwise stated.
 
 \subsection{$S'_{q=1}$ and the two-point function of $H_\tau$}
 
 We may calculate $S'_{q=1}$ using~\eqref{Expandq} and~\eqref{En}:
 \es{}{
 S'_{q = 1} = - 2 \pi^2 \langle H_\tau H_\tau \rangle_{q=1}^\text{conn} \,.
 }
 Then, using the relation between $H_\tau$ and $T_{\tau \tau}$ and translation invariance, we may write
 \es{2pt0}{
 S'_{q = 1} = - 2 \pi^2 \Vol(\mathbb{H}^{d-1} )\int_{\mathbb{H}^{d-1}} d^{d-1} x \sqrt{g} \langle T_{\tau \tau} (0) T_{\tau \tau}(x) \rangle_{q= 1}^\text{conn} \,.
 }
  For now, we assume that $T_{\tau \tau}$ is simply the conformal stress tensor, with no additional boundary terms, and we use conformal invariance to evaluate the integral above.  
  This calculation was performed in~\cite{Perlmutter:2013gua}, where it was shown that  
 \es{2pt}{
 S'_{q = 1} = - \text{Vol}\big(\mathbb{H}^{d-1} \big) { \pi^{d/2+1} \Gamma(d/2) (d-1) \over (d+1)!} C_T \,.
 }
 $C_T$ is the coefficient of the stress-tensor two-point function, whose normalization we define shortly.
 We will re-derive~\eqref{2pt} from a perspective that will be useful when calculating $S''_{q = 1}$.  Then, we will address the scalar theory, for which $T_{\tau \tau}$ has additional boundary terms. 
 
Note that in~\eqref{2pt0} we are free to choose arbitrary values for the Euclidean times of the two stress tensors.  We will fix the first stress tensor to be at $\tau_1 = 0$, and we let $\tau_2 = \tau$ be arbitrary for the second stress tensor.  We will then see explicitly that the final result does not depend on $\tau$.

 The connected correlation functions on ${\cal H}^d$ may be calculated by utilizing a conformal transformation between that space and flat $\mathbb{R}^d$~\cite{ch2}.  Writing 
 \es{}{
 ds_{\mathbb{R}^d}^2 = dt^2 + dr^2 + r^2 d\Omega_{d-2}^2 \,,
 }
 it may be verified that the coordinate transformation 
 \es{tandr}{
 t = {\sin \tau \over \cosh u + \cos \tau} \,, \qquad r = {\sinh u \over \cosh u + \cos \tau} 
 }
 conformally maps $\mathbb{R}^d$ to ${\cal H}^d$:  
 \es{}{
 ds_{\mathbb{R}^d}^2 = \Omega^2 ds_{{\cal H}^d}^2 \,, \qquad 
 \Omega = {1 \over \cosh u + \cos \tau} \,.
 }
 
 The conformal stress tensor transforms simply under the conformal transformation;
 \es{Ttrans}{
 T_{\alpha \beta}(x) = \tilde T_{\alpha \beta}(x) + S_{\alpha \beta}(x) \,, \qquad  \tilde T_{\alpha \beta}(x) \equiv  \Omega^{d-2} {dX^a \over d x^\alpha} {dX^b \over d x^\beta} T_{ab}(X) \,,
 }
 where the $x^\mu$ are coordinates on ${\cal H}^d$ and the $X^a$ are coordinates on $\mathbb{R}^d$.  The stress tensor $T_{ab}(X) $ is that in flat spacetime.  The tensor $S_{\alpha \beta}(x)$ is an anomalous term that vanishes in odd $d$, while in even $d$, $S_{\alpha \beta}(x) = \langle T_{\alpha \beta}(x) \rangle_{{\cal H}^d}$.  
 
 In both even and odd dimensions $d$, connected correlation functions of $T_{\alpha \beta}(x)$ on ${\cal H}^d$ are equal to connected correlation functions of $\tilde T_{\alpha \beta}(x)$ in flat spacetime.  This is trivial in odd $d$, since in that case the anomalous term vanishes, while in even $d$ this may be verified through a direct calculation~\cite{Perlmutter:2013gua}. 

Using~\eqref{2pt0} and~\eqref{Ttrans}, we find that
\es{Sp1}{
S'_{q = 1} &=  -{2 \pi^2 \over 2^d} \text{Vol}\big(\mathbb{H}^{d-1} \big) \text{Vol}(S^{d-2}) \int_0^\infty du \sinh^{d-2}u \left[ \left( {1 \over \cos \tau + \cosh u} \right)^{d+2} \right.\\
&\left. \left( (1 + \cos \tau \cosh u )^2 \langle T_{tt}(0) T_{tt}(t,r) \rangle_{\mathbb{R}^d} \right. \right. \\
&\left. \left.+ 2 \sin \tau \sinh u (1 + \cos \tau \cosh u) \langle T_{tt}(0) T_{tr}(t,r) \rangle_{\mathbb{R}^d}   \right.\right. \\
&\left. \left.
 + \sin^2 \tau \sinh^2 u \langle T_{tt}(0) T_{rr}(t,r) \rangle_{\mathbb{R}^d}  \right) \right] \,,
}
 where $t$ and $r$ are related to $\tau$ and $u$ through~\eqref{tandr}.
  The flat-space two-point functions of conformal stress tensors are given by~\cite{Osborn:1993cr}
\es{}{
 \langle T_{ab} (0) T_{cd}(x) \rangle = C_T {I_{ab,cd}(x) \over x^{2d}} \,,
 }
 where 
 \es{}{
 I_{ab,cd}(x) = {1 \over 2} \left( I_{ac}(x) I_{bd}(x) + I_{ad}(x) I_{bc}(x) \right) - {1 \over d} \delta_{ab} \delta_{cd} \,, \qquad I_{ac}(x) = \delta_{ac} - 2 {x_a x_c \over x^2} \,.
 }
 
 Equation~\eqref{Sp1} simplifies considerably if we take $\tau  = \pi$, for then we only need the two-point function $\langle T_{tt}(0) T_{tt}(0,r) \rangle_{\mathbb{R}^d}$.  With this specific choice of $\tau$, it is straightforward to verify~\eqref{2pt} for all dimensions $d$.  With arbitrary $\tau$, it is more illuminating to work in a specific dimension $d$, as otherwise the equations are cumbersome.  The simplest example is $d= 2$.  In this case,~\eqref{Sp1} becomes   
 \es{Sp1_2}{
 S'_{q = 1} =- C_T {\pi^2 \over 2} \log (R/ \epsilon) \int_0^\infty du &\left( {1 \over \cos \tau - \cosh u} \right)^{4} \\
 &\big( \cos(2 \tau) \cosh(2 u) - 4 \cos \tau \cosh u + 3 \big) \,.
 }
 
 It is interesting to study the behavior of the integrand above near $u = 0$.  The leading term is ${\cal O}\big( u^0\big)$, and the coefficient is proportional to $\sin(\tau/2)^{-4}$.  So long as $\tau \neq 0, 2 \pi$, the expansion near $u = 0$ is well behaved.  If $\tau = 0$ or $2\pi$, on the other-hand, then the integrand $\propto u^{-4}$ near $u = 0$, and the integral does not converge.  Restricting $\tau \neq 0, 2 \pi$, we may perform the integral in~\eqref{Sp1_2} explicitly, and we find 
 \es{Sp1_3}{
 S'_{q = 1} = -{ \pi^2 \over 3} C_T \log(R / \epsilon) \,,
 }
 independent of $\tau$, and consistent with~\eqref{2pt}.  We learn that we should be careful to avoid taking the different stress-energy tensors inside the correlators at coincident Euclidean times, otherwise we may find divergences. We will see explicitly that the same phenomenon is realized in the calculation of $S''_{q=1}$.
 
 \subsection{$S''_{q=1}$ and the three-point function of $H_\tau$}

The calculation of $S''_{q = 1}$ is more complicated than that of $S'_{q = 1}$, primarily because the three-point function of stress-energy tensors is less constrained than the two-point function.  
We make the choices $\tau_1 = 0$ and $\tau_2 = \pi$ for the first two stress tensors, but we leave $\tau_3 = \tau$ arbitrary.  Our final result for $S''_{q = 1}$ should be independent of $\tau$, which serves as a consistency check.  We may then write
\es{Sp2}{
S''_{q = 1} &=  {8 \pi^3 \over3 \, 2^d} \text{Vol}\big(\mathbb{H}^{d-1} \big) \text{Vol}(S^{d-2})\text{Vol}(S^{d-3}) \\
 &\int_0^\infty du \int_0^\infty dv \int_{0}^{\pi} d\theta \sin^{d-3} \theta \sinh^{d-2}u \sinh^{d-2} v \left[ \left( {1 \over \cos \tau + \cosh u} \right)^{d+2}  \right. \\
 &\left.   \left( {1 \over \cosh v - 1 } \right)^d \left( (1 + \cos \tau \cosh u )^2 \langle T_{tt}(0) T_{tt}(0,r') T_{tt}(t,r) \rangle_{\mathbb{R}^d}^{\rm conn}  \right.\right. \\
&\left. \left.  + 2 \sin \tau \sinh u (1 + \cos \tau \cosh u) \langle T_{tt}(0) T_{tt}(0,r') T_{tr}(t,r) \rangle_{\mathbb{R}^d}^{\rm conn}  
\right. \right. \\
&\left. \left.+ \sin^2 \tau \sinh^2 u \langle T_{tt}(0) T_{tt}(0,r') T_{rr}(t,r) \rangle_{\mathbb{R}^d}^{\rm conn}  \right) \right] \,,
}
where $t$ and $r$ are again given by~\eqref{tandr}, and $r' = \sinh v / (\cosh v - 1) $.
We begin by considering the theory in $d= 2$, where the three-point function takes a simple form.

\subsubsection{Warmup: $d = 2$}

In complex coordinates $(z, \bar z) = (t + i r, t - i r ) $, the three-point functions in $d = 2$ are given by 
\es{3pt2d}{
\langle T(0) T(z_1) T(z_2) \rangle^{\rm conn} = {c \over z_1^2 z_2^2 (z_1 - z_2)^2} \,, \qquad \langle \bar T(0) \bar T(\bar z_1) \bar T(\bar z_2) \rangle^{\rm conn} =   {c \over \bar z_1^2 \bar z_2^2 (\bar z_1 - \bar z_2)^2} \,,
}
where $c$ is the central charge.
Here, $T(z)=2\pi T_{zz}$ and $T(\bar z) = 2\pi T_{\bar z \bar z}$, while $T_{z\bar z}=0$ by conformal symmetry. We may relate the specific three-point functions appearing in~\eqref{Sp2} to the ones above through
 \es{T3d2}{
\langle T_{tt}(0) T_{tt}(0,r') T_{tt}(t,r) \rangle_{\mathbb{R}^2}^{\rm conn} &= - \langle T_{tt}(0)  T_{tt}(0,r') T_{rr}(t,r) \rangle_{\mathbb{R}^2}^{\rm conn} \\
&= { \langle T(0)  T(w) T(z) \rangle^{\rm conn} + \langle \bar T(0) \bar T(\bar w) \bar T(\bar z) \rangle^{\rm conn}   \over (2 \pi)^3}  \,, \\
\langle T_{tt}(0)  T_{tt}(0,r') T_{tr}(t,r) \rangle_{\mathbb{R}^2}^{\rm conn}  &= i  { \langle T(0) T(w) T(z) \rangle^{\rm conn} - \langle \bar T(0) \bar T(\bar w) \bar T(\bar z) \rangle^{\rm conn}   \over (2 \pi)^3} \,,
}

Using~\eqref{T3d2},~\eqref{3pt2d}, and~\eqref{Sp2}, we find
\es{Sp2_2d}{
S''_{q = 1} &= {c \log (R / \epsilon) \over 6} {1 \over 4} \int_{-\infty}^\infty du \int_{-\infty}^\infty dv \\
&{ 2 \cos \tau \big( \cosh(v-u) - \cosh(u) \big) - \cos 2 \tau \cosh(v-2 u) - \cosh v + 2   \over \cosh^2(v/2) \big( \cos \tau + \cosh(v-u) \big)^2 (\cos \tau - \cosh u )^2 } \,.
} 
The integral is convergent so long as $\tau \neq 0, \pi, 2 \pi$.  That is, we are not allowed to take coincident Euclidean times for any of the stress-energy tensors.  If $\tau = 0, 2 \pi$, the integrand is ill-behaved at $u = 0$; if $\tau = \pi$, the integral diverges along the line $v = u$.  Restricting $\tau$ to lie away from these three points, we may integrate~\eqref{Sp2_2d} exactly, and we find
\es{}{
S''_{q = 1} = {c \over 3} \log(R / \epsilon) \,.
}
This is independent of $\tau$. It is furthermore consistent with the known $d=2$ formula\footnote{Note that~\eqref{2pSq} is also consistent with~\eqref{Sp1_3} upon setting $C_T = c / (2 \pi^2)$.} 
\es{2pSq}{
S_q = {c \over 6} \left( 1+ \frac1q \right) \log(R / \epsilon)  \,.
} 
\subsubsection{$S''_{q=1}$ in general $d$}  \label{sec: 3pt}

In general dimension $d$, we may still evaluate~\eqref{Sp2} explicitly because the three-point functions of stress tensors in flat space are fixed by conformal invariance up to three theory-dependent coefficients, which we label as ${\cal A}$, ${\cal B}$, and ${\cal C}$, following the notation of \cite{Osborn:1993cr,Erdmenger:1996yc}:
\es{TTTflat}{
\langle T_{\mu\nu}(x) T_{\rho\sigma}(y) T_{\alpha\beta}(z) \rangle_{\mathbb{R}^d}^{\rm conn} = { I_{\mu\nu, \mu'\nu'}(x-z) I_{\rho\sigma, \rho'\sigma'}(y-z) \over \abs{x-z}^{2d} \abs{y-z}^{2d}}t_{\mu'\nu', \rho'\sigma', \alpha\beta}(Z) \,.
}
Here, $t_{\mu'\nu', \rho'\sigma', \alpha\beta}(Z)$ is a tensor structure depending on the coefficients ${\cal A}$, ${\cal B}$, and ${\cal C}$~\cite{Erdmenger:1996yc}, and 
\es{Zmu}{
Z_\mu = {(x-z)_\mu \over (x-z)^2} - {(y-z)_\mu \over (y-z)^2} \,.
} 
As may be seen in~\eqref{Sp2}, we need three types of three-point functions of stress tensors, differing by the stress-tensor indices.  The simplest three-point function takes the form
\es{3Ttt}{
\langle T_{tt}(x) T_{tt}(y) T_{tt}(z) \rangle_{\mathbb{R}^d}^{\rm conn} = {1 \over \abs{x-z}^{2d}\abs{y-z}^{2d}} {8({\cal A}+ {\cal C}) -  (10{\cal A} + {\cal B} + 10{\cal C})d + 4{\cal A}d^2 \over (Z^2)^{d \over 2} 4d^2} \,,
}
while the other two three-point functions are given explicitly in Appendix~\ref{sec: 3ptdetail}.

We then proceed by substituting the explicit three-point functions into~\eqref{Sp2} and performing the integrals over $u$, $v$, and $\theta$. 
While the intermediate steps involve complicated evaluations, 
the final result is a simple expression in terms of ${\cal A}$, ${\cal B}$, and ${\cal C}$:
\begin{equation}\label{s2pert}
S''_{q=1}=\frac{4 \pi^{d+1}}{3 d^3(d+2) \Gamma(d-1)} \text{Vol}  ( \mathbb{H}^{d-1})\left ( (4 d^2-10 d+8){\cal A} - d \,{\cal B} -(10d-8){\cal C} \right ) \,.
\end{equation}

\subsection{Explicit checks of $S''_{q=1}$}

In this subsection we explicitly calculate $S''_{q=1}$ in a variety of examples where $S_q$ is known for general $q$, and we compare the resulting expressions with the perturbative results of the last subsection. The same check was performed in~\cite{Perlmutter:2013gua} for $S'_{q=1}$, with agreement in all cases considered. We consider both free theories and theories with holographic duals in a variety of dimensions.  We find agreement in all examples except those involving free scalar fields.  As already mentioned, the free scalar fields are more complicated because in that case $T_{\tau \tau}$ has additional boundary contributions.  We discuss the free scalar theory in the following subsection.  

\subsubsection{Theories with gravitational duals}\label{holcheck}

Consider CFTs in arbitrary dimension $d$ that admit holographic limits with Einstein gravity duals, for simplicity. The bulk action is taken to be
\be
S = {1\over 2\ell_p^{d-1}}\int d^{d+1} x \sqrt{-g} (R+2\Lam) \,.
\ee
The 3-point function coefficients $\A,\B,\C$ for such theories at strong coupling were computed holographically in \cite{Frolov:1987dz}:
\es{}{
\A_{\rm Ein}(d) &= -{1\over 2\ell_p^{d-1}}{2d^4\Gamma(d)\over \pi^d(d-1)^3}\\%
\B_{\rm Ein}(d) &= -{1\over 2\ell_p^{d-1}}{2d^2(d^3-d^2+1)\Gamma(d)\over \pi^d(d-1)^3}\\%
\C_{\rm Ein}(d) &= -{1\over 2\ell_p^{d-1}}{d^3(2d^2-2d-1)\Gamma(d)\over 2\pi^d(d-1)^3} \,.\\
}
Substituting these expressions into~\eqref{s2pert}, one finds
\be\lab{holcft}
S''_{q=1} = \Vol(\mathbb{H}^{d-1}){1\over 2\ell_p^{d-1}}{4\pi(4d^2-8d+3)\over 3(d-1)^2} \,.
\ee

We now compare this to the non-perturbative result for $S_q$, obtained holographically in~\cite{Hung:2011nu}. Because the calculation can be mapped to a thermal free energy calculation on $\cH^d_q$, at strong coupling one can compute the free energy of the dual black hole spacetimes. These are asymptotically AdS black hole solutions of Einstein gravity with hyperbolic spatial slices and continuously tunable temperature $T^{-1}=2\pi q$. With the asymptotic AdS radius set to unity, one finds~\cite{Hung:2011nu}
\bea\lab{hol}
S_q = {\pi q\over q-1}{\rm Vol}(\IH^{d-1}){1\over \ell_p^{d-1}}\left(2-x_q^d-x_q^{d-2}\right) \,,
\eea
where $x_q$, the radial position of the horizon as a function of $q$, is defined as
\be
x_q = {1\over qd}\left(1+\sqrt{1-2dq^2+d^2q^2}\right) \,.
\ee
Taking two derivatives of \eqr{hol} at $q=1$, one recovers \eqr{holcft}.

\subsubsection{Free fields}\label{freef}
The values of $(\A,\B,\C)$ for free fields can be found in~\cite{Osborn:1993cr,Erdmenger:1996yc}. Substituting these values into \eqr{s2pert}, we obtain the following results. For Dirac fermions, we find
\es{SD3}{
\partial_q^2 S_{q=1}^D = (-2)^{-\left\lfloor \frac{d}{2}\right\rfloor } \sqrt{\pi }\Gamma \left(\frac{d}{2}\right) \dfrac{  ( d-\frac{2}{3}) }{ d (d+2) \Gamma \left(\frac{d-1}{2}\right)}
\left\{ 
\begin{array}{cc}
\pi \, , & \text{$d$ odd} \\ 
-4 \log (R/\eps) \,, & \text{$d$ even} \,,
\end{array}
\right.
}
while for complex scalars,
\es{SS3}{
\partial_q^2 \tilde S_{q=1}^{S} = (-1)^{\left\lfloor \frac{d}{2}\right\rfloor } \Gamma^2 \left(\frac{d}{2}\right)\dfrac{ 15 d^3-48 d^2+52 d-16 }{6 (d-1)^2 (d+2) \Gamma (d+1)}
\left\{ 
\begin{array}{cc}
\pi \, , & \text{$d$ odd} \\ 
-2 \log(R/\eps) \,, & \text{$d$ even} \,.
\end{array}
\right.
}
Note that we have given the scalar-field predictions above an extra tilde.  This is to distinguish these results, which come from the general formula~\eqref{s2pert}, from the results of the explicit calculations of $S_q^S$.  We need to distinguish the two results because they disagree.  As previously advertised, the disagreement is due to the fact that for the scalar theory, there are additional boundary contributions to the stress tensor that are important.

To make the discussion above more explicit, we now compare the results~\eqref{SD3} and~\eqref{SS3} to computations of $S_q$ for all $q$.
\vs
\bul {\it  Dirac fermions}
\vs

In $d=3$, we may use the results of~\cite{Klebanov:2011uf}, where it was shown that 
\es{}{
{\cal F}_q^D =2 \int_0^\infty dz \, z \coth(\pi z) \log\left( 1 + e^{-2 \pi q z} \right) + q { \zeta(3) \over  \pi^2} \,.
}
In even dimensions the computations are simpler.  In $d = 4$, one finds \cite{Dowker:2010bu}
\es{Dr}{
S^{D}_q = -{(1 + q)(7 + 37 q^2)  \over 720 q^3} \log (R/ \epsilon) \,.
}
 In Appendix~\ref{app: ferm}, we
extend the calculation to $d=6, 8$ by computing the functional determinant of the Dirac operator on $\cH^d_q$. Taking two derivatives of all of these results, we find
  \es{fermi468}{
  d=3:&\quad  {\partial_q^2 S_{q = 1}^D} =- {7 \pi^2 \over 180} \,, \\
d=4: &\quad  {\partial_q^2 S_{q = 1}^D} =-{5\over 18}\log (R/\eps) \,, \\
d=6: &\quad  {\partial_q^2 S_{q = 1}^D} ={4\over 27}\log (R/\eps) \,, \\
d=8: &\quad  {\partial_q^2 S_{q = 1}^D} =-{11\over 150}\log (R/\eps) \,.
 }
In each case, the results match \eqr{SD3}.

\vs
\bul {\it  Complex scalars}
\vs

In $d= 3$, we may again use the results of~\cite{Klebanov:2011uf}, where it shown that for complex conformal scalars
\es{ScalarKleb}{
{\cal F}_q^S = -2 \int_0^\infty dz \, z \tanh(\pi z) \log\left( 1 - e^{-2 \pi q z} \right) + q {3 \zeta(3) \over 4 \pi^2} \,.
}
In $d= 4$, similar computations give the well-known result \cite{Casini:2010kt}
\es{Sr}{
S^{S}_q = -{(1 + q)(1 + q^2)  \over 180 q^3} \log (R/ \epsilon) \,.
}
In Appendix~\ref{app: ferm} we list the known results for $S_q^S$ for complex conformal scalars in $d = 6$ and $8$. Computing their second derivatives, we find
  \es{scal468}{
d=3: &\quad  {\partial_q^2 S_{q = 1}^S} = -{2 \pi^2 \over 45}\\
d=4: &\quad  {\partial_q^2 S_{q = 1}^S} =-{1\over 9}\log (R/\eps) \,, \\
d=6: &\quad  {\partial_q^2 S_{q = 1}^S} ={1\over 54}\log (R/\eps) \,, \\
d=8: &\quad  {\partial_q^2 S_{q = 1}^S} =-{1\over 300}\log (R/\eps) \,. \\
 }
This time, in each case we find a mismatch with \eqr{SS3}:
\es{scalmis}{
d=3:&\quad  \partial_q^2 \tilde S_{q = 1}^{S}={113\over 128}\cdot \partial_q^2 S_{q = 1}^S\\
d=4:&\quad  \partial_q^2 \tilde S_{q = 1}^{S}=\frac{8}{9}\cdot \partial_q^2 S_{q = 1}^S\\
d=6:&\quad  \partial_q^2 \tilde S_{q = 1}^{S}=\frac{113}{125}\cdot \partial_q^2 S_{q = 1}^S \\
d=8:&\quad  \partial_q^2 \tilde S_{q = 1}^{S}=\frac{313}{343}\cdot \partial_q^2 S_{q = 1}^S \,.
}

\vs
\bul {\it  Maxwell field in $d = 4$}
\vs

We may also consider the Maxwell field in $d = 4$, where the theory is conformal. In this case, an explicit computation of the R\'enyi entropy gives~\cite{Fursaev:2012mp}
\es{Vr}{
S^{V}_q = -{1 + q + 31 q^2 + 91 q^3 \over 180 q^3} \log (R/ \epsilon) \,.
}
Using this result, one finds $\partial_q^2 S_{q=1}^V=-(4 / 9) \log( R/\eps)$. This matches the general formula \eqr{s2pert} upon using $(\A,\B,\C)$ for the Maxwell field~\cite{Osborn:1993cr,Erdmenger:1996yc}.

\subsection{ $S''_{q=1}$ for the free scalar field} \label{Sec: scalar}
Earlier, we explained the source of the apparent mismatch \eqr{scalmis}; our computations leading to \eqr{s2pert} used the conformal stress tensor. But as mentioned in Sec.~\ref{sec: 2}, this is incompatible with the ordinary definitions used in computing R\'enyi entropy. It is important to correctly take the boundary contributions to $T_{\tau \tau}$ into account in order to match the known results for $S''_{q=1}$. 

Before moving on to the calculation, note that $S'_{q=1}$ suffers no such ambiguity. The reason is that, as we will see below, $\langle H^{\text{conf}}_{\tau} H_{\tau} \rangle-\langle H^{\text{conf}}_{\tau} H^{\text{conf}}_{\tau} \rangle \propto\langle H^{\text{conf}}_{\tau} \int \nabla^2 \phi^2 \rangle$, which vanishes because of conformal symmetry.  Here, $H^{\text{conf}}_\tau$ uses the conformal stress tensor, while $H_\tau$ is the correct Hamiltonian that includes the boundary terms.

\subsubsection{Mapping to flat space with a conical singularity}

Instead of evaluating $\langle H_\tau H_\tau H_\tau \rangle_{q = 1}^\text{conn}$ directly, it turns out to be convenient to instead calculate $\langle H_\tau \rangle_{q}$ and then afterwards take two derivatives with respect to $q$ and evaluate the result at $q=1$.  The reason this is convenient is because conformal symmetry constrains $ \langle \phi^2  \rangle_q=a(q)$, independent of the coordinates on ${\cal H}^d_q$.  
 This then means that
\es{}{
\langle T_{\tau \tau}(x) \rangle_q=\langle T^{\text{conf}}_{\tau \tau}(x) \rangle_q-{d-2\over 4(d-1)}\nabla^2 \langle \phi^2(x) \rangle_q=\langle T^{\text{conf}}_{\tau \tau}(x) \rangle_q
} 

We now proceed by conformally mapping $\langle T_{\tau \tau}^\text{conf} \rangle_q$ to a more convenient background, where the quantity has already been computed in $d = 3$ and $d= 4$.  In particular, we utilize a mapping from ${\cal H}^d_q$ to the singular cone, ${\cal C}_q \times \mathbb{R}^{d-2}$.  Writing the metric on ${\cal H}^d_q$ in Poincar\'e coordinates as 
\es{}{
ds^2_{{\cal H}^d_q} = d \tau^2 + {dz^2 + \sum_{i=1}^{d-2} dy_i^2  \over z^2} \,,\quad \t\sim \t+2\pi q
}
we find 
\es{cone}{
ds^2_{{\cal H}^d_q} = {1 \over z^2} ds^2_{{\cal C}_q \times \mathbb{R}^{d-2}} \,, \qquad ds^2_{{\cal C}_q \times \mathbb{R}^{d-2}} = d z^2 + z^2 d \tau^2 + \sum_{i=1}^{d-2} dy_i^2 \,.
}
${\cal C}_q$ is the two-dimensional flat space with a conical singularity at the origin $z = 0$, since $\tau$ has period $2 \pi q$.

By symmetry and scaling arguments, we know that 
\es{}{
\langle T_{\tau \tau}^\text{conf}(\tau, z, y_i) \rangle^{{\cal C}_q \times \mathbb{R}^{d-2}}_q = {F(q) \over z^{d-2}} \,,
}
for some function $F(q)$ that vanishes at $q = 1$. It then follows that  
\es{Sqm}{
\left. \partial^m_q S_q \right|_{q = 1} = {2 \pi \over m + 1} \Vol(\mathbb{H}^{d-1}) \left. \partial_q^m F(q) \right|_{q = 1} \,,
}
where we have used the conformal mapping
\es{}{
\langle T_{\tau \tau}^\text{conf}(\tau, z, y_i) \rangle^{{\cal H}^d_q}_q-\langle T_{\tau \tau}^\text{conf}(\tau, z, y_i) \rangle^{{\cal H}^d}_1 = F(q) ~.
}

In $d=4$, $\langle T_{\tau \tau}^\text{conf} \rangle_q$ was computed in the background of the cone in \cite{Frolov:1987dz}, with the result
\es{Fq}{
F(q) = {(q^2-1) \big[ (q^2 -1) 3 (2 b -c) -2 a (q^2 + 3) \big] \over 23040 \pi^2 q^4} \,,
}
where 
\es{Fq1}{
a = 24 n_S + 72 n_D + 144 n_V \,, \qquad b = -8 n_S - 44 n_D - 248 n_V \,, \qquad c = -240 n_V \,,
}
and $n_S$, $n_D$, $n_V$ are the numbers of complex scalars, Dirac fermions, and vector fields, respectively.  Substituting~\eqref{Fq} into~\eqref{Sqm}, we may verify that this formula agrees with the explicit results for the free scalar, fermion, and vector field in~\eqref{Sr},~\eqref{Dr},~\eqref{Vr}, respectively.\footnote{It was recently shown in \cite{AE} that in a general $d=4$ CFT, $F(q)$ is proportional to $f_c(q)$, a function appearing in the log term of the R\'enyi entropy for generic entangling surfaces \cite{Fursaev:2013fta}.}

 For the $d=3$ scalar, the expression for $F(q)$ is somewhat more complicated than the functions found in $d = 4$.  The calculation was performed in~\cite{Souradeep:1992ia}, with the result 
 \es{Fqs3}{
 F(q) = {1 \over 8 \pi^2} \int_0^\infty du {1 \over \sinh u} \left( {\coth u \over \sinh^2 u} - {1 \over q^3} { \coth(u/q) \over \sinh^2(u/q)}  \right) \,.
 }
 Substituting~\eqref{Fqs3} into~\eqref{Sqm}, we may verify explicitly that the formula above is in agreement with the explicit computation of the R\'enyi entropy on ${\cal H}^3_q$ performed in~\cite{Klebanov:2011uf} (see the result quoted in~\eqref{ScalarKleb}).
 
The scalar-field result for $S''_{q=1}$ may also be understood by an explicit computation of $\langle H_\tau H_\tau H_\tau \rangle_{q=1}^\text{conn}$, so long as we include the correct boundary contributions to the stress tensor.  In Appendix~\ref{scalarChecks} we perform this check in two ways.  The first methods involves conformally mapping this correlation function from ${\cal H}^d$ to the sphere $S^d$.  The second method involves directly computing $\langle H_\tau H_\tau H_\tau \rangle_{q=1}^\text{conn}$ on ${\cal H}^d$.

The mapping from the ${\cal H}^d$ to $S^d$ has a subtlety that is worth some explanation.  With the inclusion of the boundary terms, the stress tensors do not transform nicely under conformal transformations.  To circumvent this fact, we use the observation that on ${\cal H}^d_q$, $\langle T_{\tau \tau} \rangle_q = \langle T_{\tau \tau}^\text{conf} \rangle_q$.  Then, we conformally map the one-point function of $T_{\tau \tau}^\text{conf}$ to the one-point function of the conformal stress tensor on the multi-covered sphere, take two derivatives with respect to $q$, which brings down two non-conformal stress tensors, and then set $q = 1$.  In the end, we see that on $S^d$ we need to evaluate the three-point function of one conformal stress tensor with two non-conformal stress tensors.  This is shown explicitly in Appendix~\ref{scalarChecks}.

To summarize, we have seen explicitly that the mismatch noted in~\eqref{scalmis} is explained by the boundary contributions to the Hamiltonian.

\subsec{R\'enyi entropy in $\N=4$ super-Yang-Mills}
As a corollary to the previous results, we are able to resolve a puzzle raised in~\cite{Perlmutter:2013gua} about the behavior of $S''_{q=1}$ in $\N=4$ SYM as a function of the `t Hooft coupling $\lambda$. The puzzle is as follows. Because $S''_{q=1}$ is fixed by the stress-tensor 3-point function on $\IR^4$, one expects
that this quantity is independent of the coupling $\lambda$: $\p_{\lam} S''_{q=1}(\lam)=0$. 
One may check this conjecture by comparing the explicit R\'enyi entropies at weak and strong coupling in the large-$N$ limit.  However, using the free field results for scalars, fermions and vectors, and comparing the result to \eqr{holcft} at $d=4$, one finds that~\cite{Perlmutter:2013gua}
\be\la{weakstrong}
\lim_{\lam\rar 0}S''_{q=1}(\lam) \neq \lim_{\lam\rar \infty}S''_{q=1}(\lam) \,.
\ee
In particular, 
\es{weakstrong2}{
\lim_{\lam\rar 0}S''_{q=1}(\lam) =-{ 4 \over 3} N^2 \log (R / \epsilon) \,, \qquad \lim_{\lam\rar \infty}S''_{q=1}(\lam) =-{35 \over 27} N^2 \log (R / \epsilon) \,.
}
 Evidently, $S''_{q=1}(\lam)$ is given by some non-trivial function of $\lam$. What failed about the non-renormalization argument?

The culprit is the contribution of the scalar field. As we have shown, one should really be using the non-conformal stress tensor in the three-point function on $\IR^4$. This object is not subject to a non-renormalization theorem. In particular, the difference between stress tensors, $T_{\t\t}^{\rm conf} - T_{\t\t} \sim \nabla^2 \phi^2$, is a conformal descendant of the Konishi operator. This operator is not protected by supersymmetry and decouples at strong coupling, where it acquires an anomalous dimension $\Delta\sim \l^{1/4}$. Therefore, the non-renormalization conjecture is incorrect, and $S''_{q=1}(\l)$ is indeed a non-trivial function of $\l$.

This is consistent with the result of Sec. \ref{holcheck}. The general lesson is that at strong coupling, the regularized and singular cones merge into one prescription, because the boundary terms in the stress tensor are suppressed. 

As a further check on this interpretation, the $\l$-dependence of $S''_{q=1}(\l)$ should not be visible at any order in perturbation theory around $\l=\infty$. We can confirm this at first non-trivial order in $\alpha'$ corrections: in \cite{Galante:2013wta}, the first correction to \eqr{holcft} in type IIB supergravity due to $O(\alpha'^3)$ corrections involving the metric and five-form was computed. The correction scales like $(q-1)^3$; this completes the argument. 

\section{Perturbative entanglement entropy and stationarity} 

The fact that the modular Hamiltonian may include boundary terms when working with the singular cone may help resolve the EE stationarity puzzle~\cite{Klebanov:2012va}.  Within the context of the $F$ theorem~\cite{Jafferis:2011zi,Klebanov:2011gs,Myers:2010xs}, stationarity is typically thought of as a requirement for the REE~\cite{Liu:2012eea} to be a well-behaved interpolating function between the $F$ values of the UV and IR fixed points along an RG flow.
More specifically, the REE is said to be stationary if it satisfies the following criteria.  Suppose we perturb away from a conformal fixed point using a relevant scalar operator ${\cal O}$, of dimension $\Delta < 3$, that lives in the UV CFT.  That is, the UV action is perturbed by $\delta I = \lambda \int d^3x {\cal O}$, where $\lambda$ is the coupling constant.  The REE ${\cal F}(g)$ is a non-trivial function of the dimensionless coupling $g = \lambda R^{3 - \Delta}$, where $R$ is the radius of the circular entangling surface.  Stationarity is the requirement that ${\cal F}'(g = 0) = 0$.

In~\cite{Klebanov:2012va,Safdi:2012sn} it was shown, through explicit lattice calculations, that the REE is not stationary for the free massive scalar RG flow, while it is stationary for the free massive fermion RG flow.  In~\cite{Nishioka:2014kpa} it was argued that in holographic RG flows, the REE is not stationary for $\Delta \leq 1$. 

We may make progress in understanding the non-stationarity by using the recently-discovered result for perturbative EE~\cite{Rosenhaus:2014woa}: under the previous perturbation $\delta I$,
\es{stat1}{
\delta S = - 2 \pi \lambda \int \langle {\cal O} H_0 \rangle^\text{conn}_{q =1} \,.
}
Above, $H_0$ is the modular Hamiltonian, with the normalization $\rho_0 = e^{-2 \pi H_0}$, where $\rho_0$ is the reduced density matrix, and $\delta S$ is the first correction to the UV EE $S_0$: $S(g) = S_0 + \delta S + O(g^2)$.  Note that it is sometimes convenient to compute $\langle {\cal O} \rangle_q$ at finite $q$ and then evaluate the first derivative with respect to $q$ instead of using~\eqref{stat1} directly:
\es{S1pt}{
\delta S =  \lambda \int \left. \partial_q \langle {\cal O} \rangle_{q} \right|_{q = 1} \,.
 }
 
The modular Hamiltonian is known explicitly for the circular entangling surface~\cite{ch2}.  It is given by 
\es{}{
H_0 = \int_0^R dr \, r \int_0^{2 \pi} d \phi { R^2 - r^2 \over 2 R} T^{00}(r, \phi) + c' \,,
}
where $c'$ is an unimportant constant.  Then, using the fact that both $\langle {\cal O} T^{00} \rangle_{q = 1}^\text{conn}$ and $\langle {\cal O} \rangle_{q = 1} $ vanish for all primary operators ${\cal O}$, it would seem that the REE is stationary for all perturbations where ${\cal O}$ is a conformal primary.  This logic is clearly incorrect, since in the free massive scalar RG flow ${\cal O} = \phi^2$ is a primary operator, and yet the REE is not stationary.  

There are, in fact, two effects that contribute to the resolution to this paradox, as we will show.  The first effect is that of the boundary terms mentioned in the previous section: the stress tensor $T^{00}$ is modified by boundary terms in the singular cone.  These boundary terms lead to a non-vanishing $\delta S$ at leading order in $\lambda = m^2/2$.  The second effect is more subtle; conformal perturbation theory about the massless fixed point is subject to IR divergences.  The IR-divergent terms need to be resummed in order to evaluate the REE at small mass.  After resumming the divergent terms, the REE may scale differently with $m^2$ than the naive perturbation theory suggests. 

This second effect is perhaps more general, in that it may apply to all perturbations with $\Delta \leq 1$.  Recall that the holographic calculations show that the REE is non-stationary when $\Delta \leq 1$~\cite{Nishioka:2014kpa}.  It is unlikely that this behavior is due to stress-tensor boundary terms, since these are expected to give subleading contributions at strong coupling.  However, it is likely that when $\Delta \leq 1$, conformal perturbation theory of the EE about the UV fixed point breaks down due to the emergence of IR divergences.
\subsection{Warmup: stationarity across the half plane} \label{HP}

While the circular entangling geometry is that of most interest, it is useful to build intuition from the calculation of the EE across the half plane. With the flat space metric written as $ds^2 = r^2 d\tau^2 + dr^2 + dx^2$, the entangling surface is simply given by $r = 0$ for all $x$.
 It is convenient to regularize the $x$ direction by taking $x \in (-L/2, L/2)$, with $L$ large.

We may calculate the EE of the free massive scalar field exactly in this geometry.  The calculation was performed in~\cite{Hertzberg:2012mn} by using an exact result for the scalar-field two-point function on the conically singular background, with the result
\es{Smhalf}{
S(m)=-\frac{L}{12} \int_{0}^{1 / \epsilon} \frac{dp}{2 \pi} \log (p^2+m^2) = - {m L \over 24}  \,.
}
The second equality above comes from removing the UV-divergent, $\epsilon$-dependent contributions to the integral.  The UV EE, in this case, is simply $S_0 = 0$, while the first correction $\delta S = - m L / 24$ is exact; that is, there are no higher-order corrections in $m$.

We would like to understand~\eqref{Smhalf} from a direct computation of the one-point function, as in~\eqref{S1pt}.  Before doing so, it is instructive to anticipate a complication that may arise, due to IR divergences in perturbation theory.  Consider expanding the integrand in~\eqref{Smhalf} around $m^2 = 0$ and then performing the integral over $p$ on each term individually.  The integrals over $p$ are IR divergent.  We may regulate the divergences with an IR cut-off $\Lambda$ in momentum space, perform the integrals, and then resum the series to recover the correct answer:
\es{Sm3}{
S(m) &= - {L \over 24 \pi} \left[ \int_\Lambda^{1 \over \epsilon} \log p^2 + \sum_{n=1}^\infty {(-1)^{n-1} \over n}   m^{2n} \int_\Lambda^{1 \over \epsilon} {1 \over  p^{2n}} \right] \\
 &= - {L \over 24 \pi} \left[ 2 \Lambda(1 - \log \Lambda) + \sum_{n=1}^\infty {(-1)^{n-1}  \over n (2n - 1)}  m^{2n} \Lambda^{1 - 2n} \right]  \\
 &= - {L \over 24 \pi} \left( 2 m \tan^{-1} {m \over \Lambda} + \Lambda\left[2 - \log(m^2 + \Lambda^2) \right] \right) \\
 &= - {L m \over 24} + O(\Lambda) \,.
}
However, if we cut off the series at any order in $m^2$, we will not obtain the correct result, even at small values of $mL$. 
Heuristically, one can think of the IR cut-off as being \mbox{$\Lambda \sim m$}.  This means that the terms $\sim$$m^{2n} \Lambda^{1 - 2n}$ may contribute to the $O(m)$ contribution to $S(m)$, even though they are naively higher order in $m^2$. This is a sign that conformal perturbation theory breaks down in this theory: it is not possible to understand the small $m$ behavior by working perturbatively around $m = 0$.

It is useful to understand this point in a more direct fashion. From~\eqref{S1pt}, we see that we need to compute the one-point function of $\phi^2$ in the conical background~\eqr{cone}. In~\cite{Souradeep:1992ia} it was shown that $\partial_q \langle \phi^2 \rangle_{q=1}=- 1 /(32 r)$. 
 Then, using~\eqref{S1pt}, we find that 
\es{deltaS2}{
\delta S = {\pi m^2 L  \over 32} \int_0^{r_\text{max}} dr =  {\pi m^2 L\, r_\text{max}  \over 32}  \,.
}  
The answer for $\delta S$ is IR-divergent; it depends on the unphysical IR cut-off $r_\text{max}$.  Heuristically, the IR cut-off $r_\text{max}$ should be of order $1/m$, since the mass $m$ is the only dynamical scale in the problem, and thus $\delta S$ is effectively of order $m$, consistent with~\eqref{Sm3}.  
Again, we see that $\delta S$ does not admit an expansion in $m^2$. To do the calculation correctly, we would need to evaluate all of the higher-order corrections to $S(m)$, in increasing powers of $m^2$, and then perform the sum over all such corrections keeping $r_\text{max}$ explicit.  

This example illustrates the fact that the conformal perturbation theory of EE breaks down for the scalar theory in $d = 3$ spacetime dimensions.  We see the same issues arising for the EE across the circular entangling surface. 

This is similar in spirit to the IR divergences often found in thermal field theory (see~\cite{le2000thermal} and~\cite{kapusta2006finite} for reviews). The presence of finite temperature can make loop diagrams IR divergent. In simple cases, one can consistently re-sum the divergent terms and have a well-defined perturbative expansion. For example in a massless $\lambda \phi^4$ theory, the one-loop re-summation of the IR divergences gives a mass to the scalar $m^2=\lambda T^2$, where $T$ is the temperature. Physical quantities, such as the pressure or the free energy, will be non-analytic functions of the coupling. It would be interesting if a similar kind of re-summation could be applied to our case. 

\subsection{Non-stationarity of the scalar field REE }

Now we consider the circular entangling surface, which is the case of most interest.  Here, of course, we do not have the exact answer for the EE, and we must rely on perturbation theory as much as we can.  The one-point function of $\phi^2$ in the conical background can be obtained by conformally mapping the Rindler results (or doing the explicit calculation, as in~\cite{Cardy:2013nua})
\es{}{
\partial_{q} \langle \phi^2(t,r,\phi) \rangle_{q=1} = - { 1 \over 32 \bar r} \,, \qquad \bar r^2 = {(r^2 + t^2 -R^2)^2 + 4 R^2 t^2 \over 4 R^2}   \,.
}
The Euclidean metric is written as $ds^2 = dt^2 + dr^2 + r^2 d \phi^2$, and the entangling surface sits at $t = 0$ and $r = R$.  The distance $\bar r$ is the conformal distance between a point in spacetime and the entangling surface.  The first correction to the EE is then given by 
\es{}{
\delta S = {\pi m^2 \over 32} \int_{-\infty}^{+\infty} dt \int_{0}^{r_\text{max}} dr {r \over \bar r} \,.
}
The integral above is not convergent; $\delta S$ diverges linearly with the IR cut-off $r_\text{max}$.  

However, we are more interested in the REE than in the EE itself.  The first correction to the REE is given by 
\es{}{
\delta {\cal F} = - \delta S + R\, \partial_R\, \delta S = -{ \pi (mR)^2 \over 8} \int_{- \infty}^{+\infty} dt \int_{0}^\infty dr {r (1 - r^2 + t^2) \over \left( r^4 +2 r^2 (t^2 -1) + (1+t^2)^2 \right)^{3/2} } \,.
} 
The integral above is now convergent, and a numerical evaluation gives\footnote{
The example in Sec.~\ref{sec: statE} can be regarded as a toy model for non-stationarity.  However, we stress that in that section we considered a mass deformation on ${\cal H}^3$, which is physically different from a mass deformation in flat space.  Thus, it is not surprising that the numerical value found in that section does not match those in this section.
}   
\es{deltaFAn}{
\delta {\cal F} \approx - 0.199221 (mR)^2 \,.
}
This result is close to the most precise numerical results from the lattice calculation of the massive REE~\cite{Nishioka:2014kpa}:
\es{}{
\delta {\cal F}^\text{lattice} \approx -0.13 (mR)^2 \,.
}
It would be nice to increase the precision of the lattice calculations in order to make a better comparison with~\eqref{deltaFAn}.

The reader should keep in mind that we do not necessarily expect~\eqref{deltaFAn} to be the final answer for the order $(mR)^2$ correction to ${\cal F}$. At order $m^2$, the IR-divergent correction to $\delta S$ was $\sim$$m^2\, R\, r_\text{max}$. Since the IR divergent term was linear in $R$, the REE had no non-trivial dependence on the IR cut-off. 
 At higher orders in perturbation theory, we do not know whether or not the IR divergences in the EE will be beyond linear order.  If there are IR divergences in the EE that are accompanied by factors $\sim$$R^n$, with $n > 1$, then the REE will be, at each order in perturbation theory, a function of the cut-off.  In that case, all corrections must be summed together to obtain an $r_\text{max}$-independent result.  

In summary, independent of whether or not the expansion is IR divergent  in this case, the boundary terms entering into the modular Hamiltonian in the singular cone give rise to non-stationary behavior. But if the expansion is IR divergent, the first-order calculation doesn't necessarily encode the whole contribution to order $(m R)^2$. IR divergences could even lead to contributions with non-analytic powers of the coupling.  

If instead we consider the REE in the regularized cone, then the first-order contribution to the REE vanishes by conformal invariance, since in that case there are no boundary contributions to the stress tensor. In this case, the leading contribution to the REE will be second-order in the coupling.  However, the REE could still suffer from IR divergences, which could give rise to non-stationary behavior.
  
 At strong coupling the regularized and singular cones yield identical results. However, it was  noted in~\cite{Nishioka:2014kpa} that, holographically, the REE is not stationary when $\Delta \leq 1$.  
We believe that this is due to IR divergences that appear in perturbation theory for $\Delta \leq 1$. 
Of course, in this section, we only studied the free scalar field.  Moreover, we could only work to first order in perturbation theory, where the IR divergences are not visible.  
To understand the non-stationarity for general perturbations, it would be useful, for example, to perform conformal perturbation theory to second order for general perturbations.  We leave this to future work.
 
 We stress, though, that when there are IR divergences, we really should not trust conformal perturbation theory at all, if we cannot re-sum the divergent terms to get an IR-cut-off-independent result.  Thus, when $\Delta \leq 1$, the more conservative conclusion is that we cannot evaluate non-stationarity based on conformal perturbation theory, because conformal perturbation theory does not make sense in this case.

\section{Conical remarks}
We have shown that accounting for subtleties of CFT on the singular cone explains puzzles in the behavior of the R\'enyi entropy and the REE for conformal scalar fields; in particular, accounting for boundary terms in the modular Hamiltonian is necessary to match previously derived results. 

Having established this, let us take a broader view on the role of conical spaces in computing entanglement. The regularized cone is, in many ways, a cleaner space in which to study CFT; for instance, it is smooth and without boundary, the stress tensor is the conformal one, and perturbative entanglement and R\'enyi entropies can be phrased in terms of conformal correlators. The regularized cone is also singled out if one wishes to preserve supersymmetry, as in the computation of super-R\'enyi entropy in $d=3$ \cite{Nishioka:2013haa}. Given the different ways of taming the conical singularity of a replicated manifold, any calculation of a given measure of entanglement, such as R\'enyi entropy, must be understood with respect to a choice of regularization. 

A natural question, distinct from that addressed so far in this paper, is whether there is a physically-preferred notion of regularization in the definition of entanglement.  
We have shown that the R\'enyi entropy of a scalar field is sensitive to the choice of regularization.  It would be interesting to compute the scalar field entropies using the regularized cone instead. 
One would also like to understand
how the different regularization methods are manifest when using lattice techniques to compute entanglement and R\'enyi entropy. This inquiry is similar in spirit to~\cite{Ohmori:2014eia}. 

While regularized cones have been used to compute EE and black hole entropy (e.g. ~\cite{Fursaev:1995ef,Fursaev:2013fta}), the same cannot yet be said about R\'enyi entropy. There has not been an understanding of the curvatures at finite $q$, or even beyond linear order around $q=1$, for example. We believe that the inherent conformal properties of regularized cones warrant further investigation of their role in understanding CFT entanglement.

\section*{Acknowledgments}

We thank Gim-Seng Ng, Janet Hung, Misha Smolkin, Markus Luty, Per Kraus, and Vladimir Rosenhaus for helpful discussions.  We are especially thankful to Igor R. Klebanov for discussions during the early stages of this project and for comments on a draft. E.P. thanks Janet Hung for early discussions on the conformal scalar discrepancy. B.R.S. and E.P. wish to thank the Aspen Center for Physics for hospitality during the final stages of this work, which was supported in part by National Science Foundation Grant No. PHYS-1066293. E.P. also wishes to thank the Michigan Center for Theoretical Physics for hospitality during this work.  B.R.S and J.L. were supported in part by the US NSF grant PHY-1314198. E.P. has  received funding from the European Research Council under the European Union's Seventh Framework Programme (FP7/2007-2013), ERC Grant agreement STG 279943, “Strongly Coupled Systems”. A.L. acknowledges support from ``Fundacion La Caixa". J.L. also received support from the Samsung Scholarship.
 
 \begin{appendix}
 
 \section{Three-point functions of stress tensors} \label{sec: 3ptdetail}

To evaluate $S''_{q=1}$, we need explicit expressions for the three-point functions of stress tensors entering into~\eqref{Sp2}.  One of these three-point functions was given in~\eqref{3Ttt}.  In this Appendix we give the other two three-point functions. Using~\eqref{TTTflat} and following the definition of $t_{\mu'\nu', \rho'\sigma', \alpha\beta}(Z)$ in~\cite{Erdmenger:1996yc}, we find
\es{2Ttt1Tti}{
\langle T_{tt}(x) T_{tt}(y) T_{ti}(z) \rangle_{\mathbb{R}^d}^{\rm conn} &= {1 \over \abs{x-z}^{2d}\abs{y-z}^{2d}} {1 \over 4d^3} {\hat{Z}_t \hat{Z}_i \over (Z^2)^{d \over 2}} \{ 2 {\cal A}(-4 + d (4 + d ((-1 + d)^2 \\
&+ (-2 + d)^2 (2 + d) \hat{Z}_i \hat{Z}_t - (-2 + d)^2 (2 + d) \hat{Z}_t^2))) \\
& + {\cal B} (4 + d (2 + d (-2 - d + 2 (-4 + d^2) \hat{Z}_i \hat{Z}_t + 4 (2 + d - d^2) \hat{Z}_t ^2 \\
& + (-2 + d) d (4 + d) \hat{Z}_t ^4))) \\
&- 2{\cal C} (8 + d (12 + d (-4 + 3 d + 4 (-4 + d^2) \hat{Z}_i \hat{Z}_t  \\
&- 2 (8 + d (-2 + 3 d)) \hat{Z}_t^2 + d (8 + d (2 + d)) \hat{Z}_t^4))) \}
}
and
\es{2Ttt1Trr}{
\langle T_{tt}(x) T_{tt}(y) T_{ij}(z) \rangle_{\mathbb{R}^d}^{\rm conn} &=   {1\over 4d^3 \abs{x-z}^{2d}\abs{y-z}^{2d} (Z^2)^{d \over 2}}  \\
&\big\{ d \hat{Z}_i \hat{Z}_j (2 {\cal A} (-8 + d^2 + (-2 + d)^2 d (2 + d) \hat{Z}_t^2) \\
&+  {\cal B} d (10 - 2 d - d^2 + (-2 + d) (4 + d (6 + d)) \hat{Z}_t^2) \\
 &+ 2 {\cal C} (-8 + d (2 + 16 \hat{Z}_t^2 + d (4 + d - (2 + d) (4 + d) \hat{Z}_t^2)))) \\
 &+\delta_{ij} \{ {\cal B} d (-2 + (-2 + d) d (1 - (-2 + d) \hat{Z}_t^2)) \\
 &+ 2 {\cal A} (8 + d (12 - 3 d - 2 d^2 + (-2 + d) (8 + d (8 + 3 d)) \hat{Z}_t^2)) \\
 &+ 2 {\cal C} (8 + d (2 + 4 d - d^2 + (-16 + d (4 + (-10 + d) d)) \hat{Z}_t^2)) \} \,,
}
where $\hat{Z}_t = {Z_t \over \abs{Z}}$, $\hat{Z}_i = {Z_i \over \abs{Z}}$.

 \section{Free-field R\'enyi entropies in even dimensions} \label{app: ferm}
  
The universal part of the R\'enyi entropy across $S^{d-2}$ in a free theory can be computed by conformally mapping the theory onto $\cH_q^d$ and computing its free energy, $\cF_q=-\log Z_q$. The $d=2$ R\'enyi entropy for a single interval is given by the well-known result \eqr{2pSq}; the $d=4$ R\'enyi entropies for free scalars, fermions and vector fields were given in Section \ref{freef}. Here, we give the results for $d=6,8$ for free complex scalars and Dirac fermions.

 \subsec{Free conformal complex scalars}
We quote the known results derived in \cite{Casini:2010kt}:
\bea\label{scalrenyi}
d=6&:&\quad S_q^S = {(1 + q) (1 + 3 q^2) (2 + 3 q^2)\over 15120 q^5}\log (R/\eps) \\
d=8&:&\quad S_q^S = -{(1 + q) (3 + 23q^2 + 79 q^4 + 79 q^6)\over 907200 q^7}\log (R/\eps)\nonumber
\eea
We have used the regulated hyperbolic volume \eqr{hypvolume}.

\subsec{Free massless Dirac fermions}
The $d=4$ result, efficiently computed using heat kernel methods, can be quickly extended to other even dimensions: 
\bea\lab{Diracrenyi}
d=6&:&\quad S_q^D={(1 + q) (31 + 276 q^2 + 1221 q^4)\over 60480 q^5}\log (R/\eps)  \\
d=8&:&\quad S_q^D=-{(1 + q) (381 + 4721 q^2 + 30103 q^4 + 
   124603 q^6)\over 14515200  q^7}\log (R/\eps) \,.  \nonumber 
\eea
The necessary ingredients for these computations can be found in e.g. \cite{Camporesi:1992tm}, \cite{Lewkowycz:2012qr}, which we briefly review now. 

We compute $\log Z_q$ of a massless Dirac fermion in even $d$ via the trace of the coincident heat kernel on $\cH_q^d$, denoted $\Tr K_{\cH_q^d}(t)$: 
\be
\log Z_q = -\half \Vol(\cH_q^d) \int_0^{\infty} {dt\over t} \Tr K_{\cH_q^d}(t) \,.
\ee
On a product space like $\cH_q^d$, $\Tr K_{\cH_q^d}(t)$ factorizes:  
\be
\Tr K_{\cH_q^d}(t) = \Tr K_{S^1_q}(t) \times \Tr K_{\IH^{d-1}}(t) \,,
\ee
where, in the coincident limit,
\bea\lab{hyphk}
\Tr K_{S^1_q}(t) &=& {2\over (4\pi t)^{1/2}} \sum_{n=1}^{\infty}(-1)^n \exp\left({-(n \pi q)^2\over t}\right)\\
\Tr K_{\IH^{d-1}}(t) &=& {1\over (4\pi t)^{1/2}}\left(-{1\over 2\pi}{\p\over \p \cosh x}\right)^{d-2\over 2} \left( \cosh ^{-1}{x\over 2}\exp \left({-x^2\over 4t}\right)\right)\Bigg|_{x=0} \,.
\eea
As is conventional, we have subtracted from $\Tr K_{S^1_q}(t)$ the $n=0$ zero mode contribution. The trace over spinor indices is left implicit so far. One can write $\Tr K_{\IH^{d-1}}(t) $ as
\be
\Tr K_{\IH^{d-1}}(t)  = (4\pi t)^{-{d-1\over 2}}\sum_{n=0}^{{d-2\over 2}}c_nt^{n}
\ee
for some constants $c_n$ defined by matching to \eqr{hyphk}. (It is straightforward to show that $c_0=1$.) Combining the previous several equations, and restoring an overall $2^{d/2}$ from the trace over spinor indices, one finds
\be\lab{logz2}
\log Z_q =-2^{d+2\over 2}\pi q \Vol(\IH^{d-1})\int_0^{\infty} {dt\over t} (4\pi t)^{-d/2}\sum_{m=1}^{\infty}(-1)^m \sum_{n=0}^{{d-2\over 2}} c_n t^{n} \exp\left(-{(m \pi q)^2\over t}\right) \,.
\ee
Using the Riemann zeta function relation
\be
\sum_{m=1}^{\infty} (-1)^m m^{-s} = (2^{1-s}-1)\zeta(s) 
\ee
and an integral representation of the Gamma function, one can evaluate \eqr{logz2} in zeta function regularization as
\be\label{logZ3}
\log Z_q =-\left(2\pi\right)^{-{d-2\over 2}}q\Vol(\IH^{d-1}) \sum_{n=0}^{{d-2\over 2}} c_n (q\pi)^{2n-d}\Gamma({d\over 2}-n)(2^{1-d+2n}-1)\zeta(d-2n) \,.
\ee

In $d=4,6,8$, the constants $c_n$ are
\bea
d=4&:&\quad c_0 = 1~,\quad c_1=\half \nonumber\\
d=6&:&\quad c_0 = 1~,\quad c_1={5\over 3}~, \quad c_2 = {3\over 4}\\
d=8&:&\quad  c_0 = 1~,\quad c_1={7\over 2}~, \quad c_2 = {259\over 60}~, \quad c_3 = {15\over 8} \,.\nonumber
\eea
Substituting these into \eqr{logZ3} and using the definition of R\'enyi entropy yields the desired result \eqr{Diracrenyi}.

\section{Additional checks for the scalar $S''_{q=1}$} \label{scalarChecks}

In this Appendix we explicitly calculate $\langle H_\tau H_\tau H_\tau \rangle_{q = 1}^\text{conn}$ for the scalar field, including the correct boundary terms.   We verify that once the boundary terms are included, the calculations agree with the explicit results for $S''_{q=1}$.

\subsection{The three-point function on the sphere} \label{3pt:sphere}

The calculation on the sphere utilizes the following conformal transformation from ${\cal H}^d$ to $S^d$:
\es{spheremetric}{
ds^2_{{\cal H}^d} &= d\tau^2 + d\rho^2 + \sinh^2\rho d\Omega^2_{d-2} \\
&= {1 \over \sin^2\theta} (\sin ^2 \theta d\tau^2 + d\theta^2 + \cos^2\theta d\Omega^2_{d-2}) = {1\over \Omega^2}ds^2_{S^d} \,.
} 
where $\sinh\rho=\cot\theta$. Under the conformal transformation, the  scalar propagator transforms by
\es{sphereprop}{
\langle \phi(x) \phi(y) \rangle_{S^d} = \Omega(x)^{d-2 \over 2}\Omega(y)^{d-2 \over 2} \langle \phi(x) \phi(y) \rangle_{{\cal H}^d} \,.
}
Now, we may calculate the three-point function of modular Hamiltonians 
\es{Hamiltonian3pt}{
\langle H_\tau H_\tau H_\tau \rangle = \int d^{d-1}x_1 \sqrt{g_1} \int d^{d-1}x_2 \sqrt{g_2} \int d^{d-1}x_3 \sqrt{g_3} \langle T_{\tau\tau}(x_1)T_{\tau\tau}(x_2)T_{\tau\tau}(x_3) \rangle \,,
}
with the stress tensors $T_{\mu\nu}$ given by
\es{Talpha}{
T^\alpha_{\mu\nu} = \partial_\mu \phi \partial_\nu \phi - {1 \over 2}g_{\mu\nu} (\partial \phi)^2 + \xi (\alpha {\cal R}_{\mu\nu} - {1 \over 2} g_{\mu\nu} {\cal R}) \phi^2 - \alpha \xi (\nabla_\mu \nabla_\nu - g_{\mu\nu} \nabla^2)\phi^2 \,.
}
where $\xi = {d-2 \over 4(d-1)}$, and $\alpha=1$ (conformal) or $\alpha=0$ (non-conformal). For this section only, we have introduced a new notation, whereby $T^{\text{conf}}_{\tau\tau} \equiv T^1_{\tau\tau}$ and $T_{\tau\tau} \equiv T^0_{\tau\tau}$, and correspondingly for $H^1_\tau$ and $H^0_\tau$.

To calculate the three-point function $\langle T^{\alpha_1}_{\tau\tau}(x_1)T^{\alpha_2}_{\tau\tau}(x_2)T^{\alpha_3}_{\tau\tau}(x_3) \rangle^\text{conn}$, it is convenient to first compute 
$G(x_1 - x_4)G(x_2 - x_5)G(x_3 - x_6)$.  Then one may take derivatives of this combination of Green's functions, along with taking appropriate limits for the coordinates, to recover the three-point function of stress tensors.

We are allowed to set $x_1 = 0$ in the correlation function and factor out a regulated volume of the hyperbolic space, $\Vol( \mathbb{H}^{d-1})$. It is then a straightforward if tedious exercise to compute the three-point function:
\es{T100}{
\langle H^1_\tau H^0_\tau H^0_\tau \rangle = 
\left\{ 
\begin{array}{cc}
-{1 \over 60 \pi} \, , & \text{$d$ = 3} \\ 
-{1 \over 24\pi^3} \log(R/\epsilon)  \, , & \text{$d$ = 4} \,,
\end{array}
\right.
}
\es{T111}{
\langle H^1_\tau H^1_\tau H^1_\tau \rangle = 
\left\{ 
\begin{array}{cc}
-{1 \over 60 \pi}{113 \over 128} \, , & \text{$d$ = 3} \\ 
-{1 \over 24\pi^3}{8 \over 9} \log(R/\epsilon)  \, , & \text{$d$ = 4} \,.
\end{array}
\right.
}
Here, $H^1_\tau$ denotes the modular Hamiltonian with the conformal stress tensor on the sphere, while $H^0_\tau$ denotes the non-conformal stress tensor. Using~\eqref{En}, we may obtain the second derivatives of the R\'enyi entropy at $q=1$. 

The three-point function of conformal stress tensors in~\eqref{T111} agrees with the results in~\eqref{scalmis}.  However, as discussed in Sec.~\ref{Sec: scalar}, the correct three-point function to compute for reproducing the $S''_{q=1}$ is that in~\eqref{T100}. This was explained in Section 3.4. Indeed, the results in~\eqref{T100} are consistent with the explicit calculations of $S''_{q=1}$ given in~\eqref{scal468}.

\subsection{The three-point function on $S^1 \times \mathbb{ H}^{d-1}$} \label{3pt:hyperboloid}

Now we will calculate the three-point functions directly on ${\cal H}^d$. In this case, we need to consider additional boundary terms at infinity that are needed to have a well-defined variational principle. That is, we need to add the boundary action $S_{\text{bdry}} = {1\over2}\int_{\rho=\infty} \sqrt{g} \delta \phi \partial_\rho \phi$ so that $\delta S + \delta S_{\text{bdry}} = 0$, when evaluated on solutions to the equations of motion. Therefore, the non-conformal stress tensor becomes
\es{unimprovedT}{
\int T_{\tau\tau} &= \int \left( (\partial_\tau \phi)^2 - {1\over 2}\partial_{\mu} \phi\partial^{\mu}\phi + {(d - 2) \over 8 (d-1)} {\cal R}   \phi^2  \right)+ \int_{\rho=\infty}{1 \over 2} \phi \partial_\rho \phi \\
&= \tilde{H}^0_{\tau} + B \,,
}
where $\tilde{H}^0_{\tau}$ refers to the modular Hamiltonian formed with the usual non-conformal stress tensor and $B$ is the additional boundary term needed on $\cH^d$. 

While the three-point function $\langle H^1_\tau H^1_\tau H^1_\tau \rangle$ doesn't receive corrections from the boundary term $B$, the three-point function $\langle H^1_\tau H^0_\tau H^0_\tau \rangle$ does receive important corrections. The necessary calculation is
\es{T111H}{
\langle H^1_\tau H^0_\tau H^0_\tau \rangle = \langle H^1_\tau \tilde{H}^0_{\tau}  \tilde{H}^0_{\tau} \rangle + 2 \langle H^1_\tau  \tilde{H}^0_{\tau} B \rangle+ \langle H^1_\tau B B \rangle \,.
}

In $d=3$, an explicit calculation gives $\langle H^1_\tau  \tilde{H}^0_{\tau}  \tilde{H}^0_{\tau} \rangle = - 17/(1920 \pi)$, $\langle H^1_\tau  \tilde{H}^0_{\tau} B \rangle = - 1 /( 128 \pi)$ and $\langle H^1_\tau B B \rangle = 1 /(128 \pi)$. In $d=4$, we find $\langle H^1_\tau  \tilde{H}^0_{\tau}  \tilde{H}^0_{\tau} \rangle = 0=\langle H^1_\tau  \tilde{H}^0_{\tau} B \rangle $ and $\langle H^1_\tau B B \rangle = - 1/ (24 \pi^3) \log(R / \epsilon)$. Substituting these results into~\eqref{T111H}, we recover the expected results given in~\eqref{T111}.

\end{appendix}

\bibliographystyle{ssg}
\bibliography{draft}

\begingroup\raggedright\begin{thebibliography}{10}

\bibitem{renyi0}
A.~R\'enyi, ``On measures of information and entropy,'' in {\em Proceedings of
  the 4th Berkeley Symposium on Mathematics, Statistics and Probability},
  vol.~1, (Berkeley, CA), p.~547, U. of California Press, 1961.

\bibitem{renyi1}
A.~R\'enyi, ``On the foundations of information theory,'' {\em Rev. Int. Stat.
  Inst.} {\bf 33} (1965), no.~1.

\bibitem{cardyCFT1}
P.~Calabrese and J.~L. Cardy, ``{Entanglement entropy and quantum field theory:
  A Non-technical introduction},'' {\em Int.J.Quant.Inf.} {\bf 4} (2006) 429,
  \href{http://xxx.lanl.gov/abs/quant-ph/0505193}{{\tt quant-ph/0505193}}.

\bibitem{Eisert:2008ur}
J.~Eisert, M.~Cramer, and M.~Plenio, ``{Area laws for the entanglement entropy
  - a review},'' {\em Rev.Mod.Phys.} {\bf 82} (2010) 277--306,
  \href{http://xxx.lanl.gov/abs/0808.3773}{{\tt 0808.3773}}.

\bibitem{Nishioka:2009un}
T.~Nishioka, S.~Ryu, and T.~Takayanagi, ``{Holographic entanglement entropy: An
  Overview},'' {\em J.Phys.A} {\bf A42} (2009) 504008,
  \href{http://xxx.lanl.gov/abs/0905.0932}{{\tt 0905.0932}}.

\bibitem{Jafferis:2011zi}
D.~L. Jafferis, I.~R. Klebanov, S.~S. Pufu, and B.~R. Safdi, ``{Towards the
  $F$-Theorem: ${\cal N}=2$ Field theories on the three-sphere},'' {\em JHEP}
  {\bf 1106} (2011) 102, \href{http://xxx.lanl.gov/abs/1103.1181}{{\tt
  1103.1181}}.

\bibitem{Klebanov:2011gs}
I.~Klebanov, S.~Pufu, and B.~Safdi, ``{$F$-Theorem without supersymmetry},''
  {\em JHEP} {\bf 1110} (2011) 038,
  \href{http://xxx.lanl.gov/abs/1105.4598}{{\tt 1105.4598}}.

\bibitem{Myers:2010xs}
R.~C. Myers and A.~Sinha, ``{Seeing a c-theorem with holography},'' {\em
  Phys.Rev.} {\bf D82} (2010) 046006,
  \href{http://xxx.lanl.gov/abs/1006.1263}{{\tt 1006.1263}}.

\bibitem{Casini:2012ei}
H.~Casini and M.~Huerta, ``{On the RG running of the entanglement entropy of a
  circle},'' \href{http://xxx.lanl.gov/abs/1202.5650}{{\tt 1202.5650}}.

\bibitem{Liu:2012eea}
H.~Liu and M.~Mezei, ``{A Refinement of entanglement entropy and the number of
  degrees of freedom},'' {\em JHEP} {\bf 1304} (2013) 162,
  \href{http://xxx.lanl.gov/abs/1202.2070}{{\tt 1202.2070}}.

\bibitem{ch2}
H.~Casini, M.~Huerta, and R.~C. Myers, ``{Towards a derivation of holographic
  entanglement entropy},'' {\em JHEP} {\bf 1105} (2011) 036,
  \href{http://xxx.lanl.gov/abs/1102.0440}{{\tt 1102.0440}}.

\bibitem{Zamolodchikov:1986gt}
A.~B. Zamolodchikov, ``{Irreversibility of the Flux of the Renormalization
  Group in a 2D Field Theory},'' {\em JETP Lett.} {\bf 43} (1986) 730--732.

\bibitem{Cardy:1988cwa}
J.~L. Cardy, ``{Is there a c-theorem in four dimensions?},'' {\em Phys.Lett.}
  {\bf B215} (1988) 749--752.

\bibitem{Komargodski:2011vj}
Z.~Komargodski and A.~Schwimmer, ``{On Renormalization group flows in four
  dimensions},'' \href{http://xxx.lanl.gov/abs/1107.3987}{{\tt 1107.3987}}.

\bibitem{Perlmutter:2013gua}
E.~Perlmutter, ``{A universal feature of CFT R{\'e}nyi entropy},'' {\em JHEP}
  {\bf 1403} (2014) 117, \href{http://xxx.lanl.gov/abs/1308.1083}{{\tt
  1308.1083}}.

\bibitem{Klebanov:2012va}
I.~R. Klebanov, T.~Nishioka, S.~S. Pufu, and B.~R. Safdi, ``{Is Renormalized
  Entanglement Entropy Stationary at RG Fixed Points?},'' {\em JHEP} {\bf 1210}
  (2012) 058, \href{http://xxx.lanl.gov/abs/1207.3360}{{\tt 1207.3360}}.

\bibitem{Nishioka:2014kpa}
T.~Nishioka, ``{Relevant Perturbation of Entanglement Entropy and
  Stationarity},'' \href{http://xxx.lanl.gov/abs/1405.3650}{{\tt 1405.3650}}.

\bibitem{Hung:2014npa}
L.-Y. Hung, R.~C. Myers, and M.~Smolkin, ``{Twist operators in higher
  dimensions},'' \href{http://xxx.lanl.gov/abs/1407.6429}{{\tt 1407.6429}}.

\bibitem{Hung:2011nu}
L.-Y. Hung, R.~C. Myers, M.~Smolkin, and A.~Yale, ``{Holographic Calculations
  of Renyi Entropy},'' {\em JHEP} {\bf 1112} (2011) 047,
  \href{http://xxx.lanl.gov/abs/1110.1084}{{\tt 1110.1084}}.

\bibitem{Casini:2010kt}
H.~Casini and M.~Huerta, ``{Entanglement entropy for the $n$-sphere},'' {\em
  Phys.Lett.} {\bf B694} (2010) 167--171,
  \href{http://xxx.lanl.gov/abs/1007.1813}{{\tt 1007.1813}}.

\bibitem{Diaz:2007an}
D.~E. Diaz and H.~Dorn, ``{Partition functions and double-trace deformations in
  AdS/CFT},'' {\em JHEP} {\bf 05} (2007) 046,
  \href{http://xxx.lanl.gov/abs/hep-th/0702163}{{\tt hep-th/0702163}}.

\bibitem{Callan199455}
C.~Callan and F.~Wilczek, ``On geometric entropy,'' {\em Physics Letters B}
  {\bf 333} (1994), no.~1--2 55 -- 61.

\bibitem{Casini:2013rba}
H.~Casini, M.~Huerta, and J.~A. Rosabal, ``{Remarks on entanglement entropy for
  gauge fields},'' {\em Phys.Rev.} {\bf D89} (2014) 085012,
  \href{http://xxx.lanl.gov/abs/1312.1183}{{\tt 1312.1183}}.

\bibitem{Fursaev:1995ef}
D.~V. Fursaev and S.~N. Solodukhin, ``{On the description of the Riemannian
  geometry in the presence of conical defects},'' {\em Phys.Rev.} {\bf D52}
  (1995) 2133--2143, \href{http://xxx.lanl.gov/abs/hep-th/9501127}{{\tt
  hep-th/9501127}}.

\bibitem{Lewkowycz:2013laa}
A.~Lewkowycz and J.~Maldacena, ``{Exact results for the entanglement entropy
  and the energy radiated by a quark},'' {\em JHEP} {\bf 1405} (2014) 025,
  \href{http://xxx.lanl.gov/abs/1312.5682}{{\tt 1312.5682}}.

\bibitem{Kabat:1995eq}
D.~N. Kabat, ``{Black hole entropy and entropy of entanglement},'' {\em
  Nucl.Phys.} {\bf B453} (1995) 281--299,
  \href{http://xxx.lanl.gov/abs/hep-th/9503016}{{\tt hep-th/9503016}}.

\bibitem{Gibbons:1976ue}
G.~Gibbons and S.~Hawking, ``{Action Integrals and Partition Functions in
  Quantum Gravity},'' {\em Phys.Rev.} {\bf D15} (1977) 2752--2756.

\bibitem{Ryu:2006ef}
S.~Ryu and T.~Takayanagi, ``{Aspects of Holographic Entanglement Entropy},''
  {\em JHEP} {\bf 0608} (2006) 045,
  \href{http://xxx.lanl.gov/abs/hep-th/0605073}{{\tt hep-th/0605073}}.

\bibitem{Faulkner:2013ana}
T.~Faulkner, A.~Lewkowycz, and J.~Maldacena, ``{Quantum corrections to
  holographic entanglement entropy},''
  \href{http://xxx.lanl.gov/abs/1307.2892}{{\tt 1307.2892}}.

\bibitem{Rosenhaus:2014woa}
V.~Rosenhaus and M.~Smolkin, ``{Entanglement Entropy: A Perturbative
  Calculation},'' \href{http://xxx.lanl.gov/abs/1403.3733}{{\tt 1403.3733}}.

\bibitem{Klebanov:2011uf}
I.~R. Klebanov, S.~S. Pufu, S.~Sachdev, and B.~R. Safdi, ``{Renyi Entropies for
  Free Field Theories},'' {\em JHEP} {\bf 1204} (2012) 074,
  \href{http://xxx.lanl.gov/abs/1111.6290}{{\tt 1111.6290}}.

\bibitem{Cardy:2013nua}
J.~Cardy, ``{Some results on the mutual information of disjoint regions in
  higher dimensions},'' {\em J.Phys.} {\bf A46} (2013) 285402,
  \href{http://xxx.lanl.gov/abs/1304.7985}{{\tt 1304.7985}}.

\bibitem{Osborn:1993cr}
H.~Osborn and A.~Petkou, ``{Implications of conformal invariance in field
  theories for general dimensions},'' {\em Annals Phys.} {\bf 231} (1994)
  311--362, \href{http://xxx.lanl.gov/abs/hep-th/9307010}{{\tt
  hep-th/9307010}}.

\bibitem{Erdmenger:1996yc}
J.~Erdmenger and H.~Osborn, ``{Conserved currents and the energy momentum
  tensor in conformally invariant theories for general dimensions},'' {\em
  Nucl.Phys.} {\bf B483} (1997) 431--474,
  \href{http://xxx.lanl.gov/abs/hep-th/9605009}{{\tt hep-th/9605009}}.

\bibitem{Frolov:1987dz}
V.~P. Frolov and E.~Serebryanyi, ``{Vacuum Polarization in the Gravitational
  Field of a Cosmic String},'' {\em Phys.Rev.} {\bf D35} (1987) 3779--3782.

\bibitem{Dowker:2010bu}
J.~Dowker, ``{Entanglement entropy for even spheres},''
  \href{http://xxx.lanl.gov/abs/1009.3854}{{\tt 1009.3854}}.

\bibitem{Fursaev:2012mp}
D.~V. Fursaev, ``{Entanglement Renyi Entropies in Conformal Field Theories and
  Holography},'' \href{http://xxx.lanl.gov/abs/1201.1702}{{\tt 1201.1702}}.

\bibitem{AE}
A.~Lewkowycz and E.~Perlmutter, ``{Universality in the geometric dependence of
  R\'enyi entropy},'' {\em to appear}.

\bibitem{Fursaev:2013fta}
D.~V. Fursaev, A.~Patrushev, and S.~N. Solodukhin, ``{Distributional Geometry
  of Squashed Cones},'' \href{http://xxx.lanl.gov/abs/1306.4000}{{\tt
  1306.4000}}.

\bibitem{Souradeep:1992ia}
T.~Souradeep and V.~Sahni, ``{Quantum effects near a point mass in
  (2+1)-Dimensional gravity},'' {\em Phys.Rev.} {\bf D46} (1992) 1616--1633,
  \href{http://xxx.lanl.gov/abs/hep-ph/9208219}{{\tt hep-ph/9208219}}.

\bibitem{Galante:2013wta}
D.~A. Galante and R.~C. Myers, ``{Holographic Renyi entropies at finite
  coupling},'' {\em JHEP} {\bf 1308} (2013) 063,
  \href{http://xxx.lanl.gov/abs/1305.7191}{{\tt 1305.7191}}.

\bibitem{Safdi:2012sn}
B.~R. Safdi, ``{Exact and Numerical Results on Entanglement Entropy in
  (5+1)-Dimensional CFT},'' {\em JHEP} {\bf 1212} (2012) 005,
  \href{http://xxx.lanl.gov/abs/1206.5025}{{\tt 1206.5025}}.

\bibitem{Hertzberg:2012mn}
M.~P. Hertzberg, ``{Entanglement Entropy in Scalar Field Theory},'' {\em
  J.Phys.} {\bf A46} (2013) 015402,
  \href{http://xxx.lanl.gov/abs/1209.4646}{{\tt 1209.4646}}.

\bibitem{le2000thermal}
M.~Le~Bellac, {\em Thermal field theory}.
\newblock Cambridge University Press, 2000.

\bibitem{kapusta2006finite}
J.~I. Kapusta and C.~Gale, ``Finite-temperature field theory: Principles and
  applications,'' {\em Cambridge Monographs on Mathematical Physics. Cambridge:
  Cambridge University Press, 2006.} {\bf 1} (2006).

\bibitem{Nishioka:2013haa}
T.~Nishioka and I.~Yaakov, ``{Supersymmetric R{\'e}nyi Entropy},'' {\em JHEP}
  {\bf 1310} (2013) 155, \href{http://xxx.lanl.gov/abs/1306.2958}{{\tt
  1306.2958}}.

\bibitem{Ohmori:2014eia}
K.~Ohmori and Y.~Tachikawa, ``{Physics at the entangling surface},''
  \href{http://xxx.lanl.gov/abs/1406.4167}{{\tt 1406.4167}}.

\bibitem{Camporesi:1992tm}
R.~Camporesi, ``{The Spinor heat kernel in maximally symmetric spaces},'' {\em
  Commun.Math.Phys.} {\bf 148} (1992) 283--308.

\bibitem{Lewkowycz:2012qr}
A.~Lewkowycz, R.~C. Myers, and M.~Smolkin, ``{Observations on entanglement
  entropy in massive QFT's},'' {\em JHEP} {\bf 1304} (2013) 017,
  \href{http://xxx.lanl.gov/abs/1210.6858}{{\tt 1210.6858}}.

\end{thebibliography}\endgroup

\end{document}